\documentclass[aps,floats,prx,showpacs,twocolumn]{revtex4-1}

\usepackage{amsfonts,amsmath} \usepackage{bm} \usepackage{dcolumn}
\usepackage{epsfig} \usepackage{latexsym}
\usepackage{graphicx}
\usepackage{multirow}

\begin{document}

\title{Charge modulation as fingerprints of phase-string triggered interference}

\author{Zheng Zhu,$^{1}$ Chushun Tian,$^{1}$ Hong-Chen Jiang,$^{2}$ Yang Qi,$^{1}$ Zheng-Yu Weng,$^{1,3}$ and Jan Zaanen$^{4}$}

\affiliation{$^{1}$Institute for Advanced Study, Tsinghua University, Beijing, 100084, China\\
$^2$Stanford Institute for Materials and Energy Sciences, SLAC National Accelerator Laboratory, 2575 Sand Hill Road,  Menlo Park, CA 94025, USA\\
$^3$ Collaborative Innovation Center of Quantum Matter, Tsinghua University, Beijing, 100084, China\\
$^4$The Institute Lorentz for Theoretical Physics, Leiden University, Leiden, The Netherlands}

\date{\today}

\begin{abstract}

Charge order appears to be an ubiquitous phenomenon in doped Mott insulators, which is currently under intense experimental and theoretical investigations
particularly in the high $T_c$ cuprates. This phenomenon is conventionally understood in terms of Hartree-Fock type mean field theory. Here we demonstrate a mechanism
for charge  modulation  which is rooted in the many-particle quantum physics arising in the strong coupling limit. Specifically, we consider the problem of a single hole in a
bipartite   $t-J$ ladder. As a remnant of the fermion signs, the hopping hole picks up subtle phases pending the fluctuating spins, the so-called phase string effect.  We demonstrate
the presence of charge modulations in the density matrix renormalization group solutions which disappear when the phase strings are switched off.  This form of
charge modulation can be understood analytically in a path-integral language { with a mean-field like approximation adopted}, showing that the phase strings give rise to constructive interferences leading to self-localization. When the latter occurs, left- and right-moving propagating modes emerge inside the localization volume and their interference is responsible for
the real space charge modulation.

\end{abstract}


\maketitle

\section{Introduction}
\label{sec:introduction}

The most ubiquitous form of order is the one involving the breaking of space translations and rotations
leading to the solid form of matter. Dealing with the solids  of everyday life, but also with the Wigner crystals
formed in low density electron systems, this is easily understood in terms of minimization of the potential energy
associated with the interactions between the constituents. Since the 1950's it has been well understood that the order can also occur
in highly itinerant systems in the form of the Peierls mechanism. This relies on the Hartree-Fock type mean field theory.
The order parameter turns into a potential diffracting the electron waves of the nearly free system;
the order is stabilized by the energy gain associated with opening a gap at the Fermi energy.

The Hartree-Fock type theory is controlled at zero temperature by the diminishing of the collective quantum fluctuations of the order and it can therefore
also be reliable when the interactions become strong. In the late 1980's it was discovered that according to
the Hartree-Fock theory the electronic stripes should be formed in strongly coupled doped Mott insulators \cite{Zaanen89,Machida89}.
This refers to textures formed from antiferromagnetic Mott-insulating domains, separated
by lines of charge (in two dimensions) which are at the same time domain walls in the spin background. Such insulating
stripes turned out to be ubiquitous in generic doped Mott insulators (nickelates, cobaltates, manganites). In 1995 a stripe-like
ordering phenomenon was discovered in the 214 family of cuprate superconductors \cite{Uchida95}, but it became immediately
clear that these were in crucial respects different from the Hartree-Fock variant: these turned out to be ``half-filled" and associated
with metallic and even superconducting states \cite{Zaanen10,Zaanen14}.  Quite recently there has been a resurgence of interest in this subject by the discovery of
``stripe like" order in the 2212 and 123 families of superconductors, which is yet behaving differently from the 214 stripes \cite{Zaanen14}.
Despite the large body of theoretical proposals (e.g., Refs.~\cite{Sachdev14d,Efetov14a,Mei14}), it appears that these are yet far from being
completely understood.

The next surprise happened in 1998, by the discovery of White and Scalapino that the density matrix renormalization group (DMRG) computations on the $t-J$ model
revealed stripes that appear to be quite literally like the ones experimentally observed in the 214 cuprates \cite{White98}.  The very recent results obtained
using the fanciful infinite projected entangled-pair state
method add further credibility to this claim \cite{Troyer14}. All along it has been obvious that the differences with the
Hartree-Fock stripes are caused by the strong quantum nature of the spin one-half system. However, up to the present day an explanation of these numerical
results in terms of a general physics principle has been lacking. Leaving a detailed analysis of this many-hole problem to future work, we focus
here on this physics principle itself in the simplest possible setting, namely, the $t-J$ model doped with a single hole.

This single hole problem has itself a long history \cite{Khomskii68,Brinkman70}. In the late 1980's it was asserted that this could be
solved using the linear spin wave--self-consistent Born approximation approach \cite{SCBA}.
One assumes that the spin system can be parametrized in terms of the spin waves, and the hole-spin wave scattering is treated in the rainbow
approximation. The outcome is an electron-like quasiparticle that propagates facilitated by the quantum spin fluctuations governed by the superexchange
interaction. However, there is a
subtlety that is ignored in this approach.  Upon hopping in the quantum spin background the hole acquires a phase of $\pi$ whenever
it exchanges its position with a down($\downarrow$)-spin, the so-called ``phase strings'' \cite{Sheng1996,Weng1997,weng_07}.
These are genuine quantum phases, that can be understood as
the ``left-overs" of the fermion signs after Mott-projection, becoming alive when mobile holes are present \cite{Zaanen08,Zaanen11}.
These give rise to interference effects which are very different from the ones familiar from simple quantum mechanics, since they are inherently tied to the quantum many-body
nature of the spin system.  It was recently \cite{ZZ2013} demonstrated that this goes so far that even in the perfect translationally invariant case the hole is subjected
to self-localization \cite{Weng2001}. In section  ~\ref{sec:interference} we will unveil the nature of the mechanism leading to this self-localization,
resting on a general  path-integral consideration { and a mean-field like approximation}. It turns out to be a close cousin of the usual mechanism leading to Anderson localization \cite{Lee85}; the crucial difference
is that it acts out in the presence of a perfect lattice translational symmetry, while the interference mechanism giving rise to enhanced backscattering can be directly traced to
the phase strings.  Along the same lines we will also demonstrate that this phase-string interference drives the spontaneous charge modulation in the single hole case,
representing a charge ordering mechanism which is uniquely tied to the many-body physics
realized in strongly interacting doped Mott insulators.

\begin{figure}[tbp]
\centerline{\includegraphics[height=1in,width=3.0in] {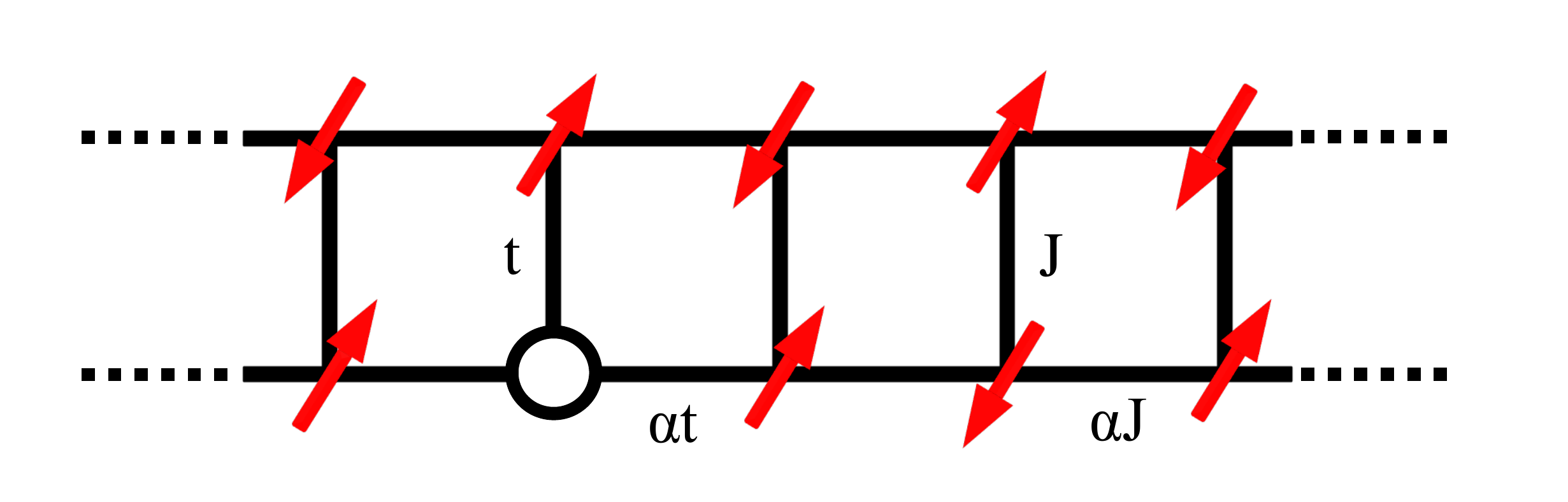}}
\caption{(Color online) The single-hole (empty circle) doped two-leg $t-J$ ladder.
The interchain hopping and superexchange parameters are $t$ and $J$, respectively, while their intrachain counterparts are $\alpha t$ and $\alpha J$, $\alpha>0$.
}
\label{Fig:Model}
\end{figure}
\begin{figure*}[tbp]
\centerline{
    \includegraphics[height=6in,width=6.2in] {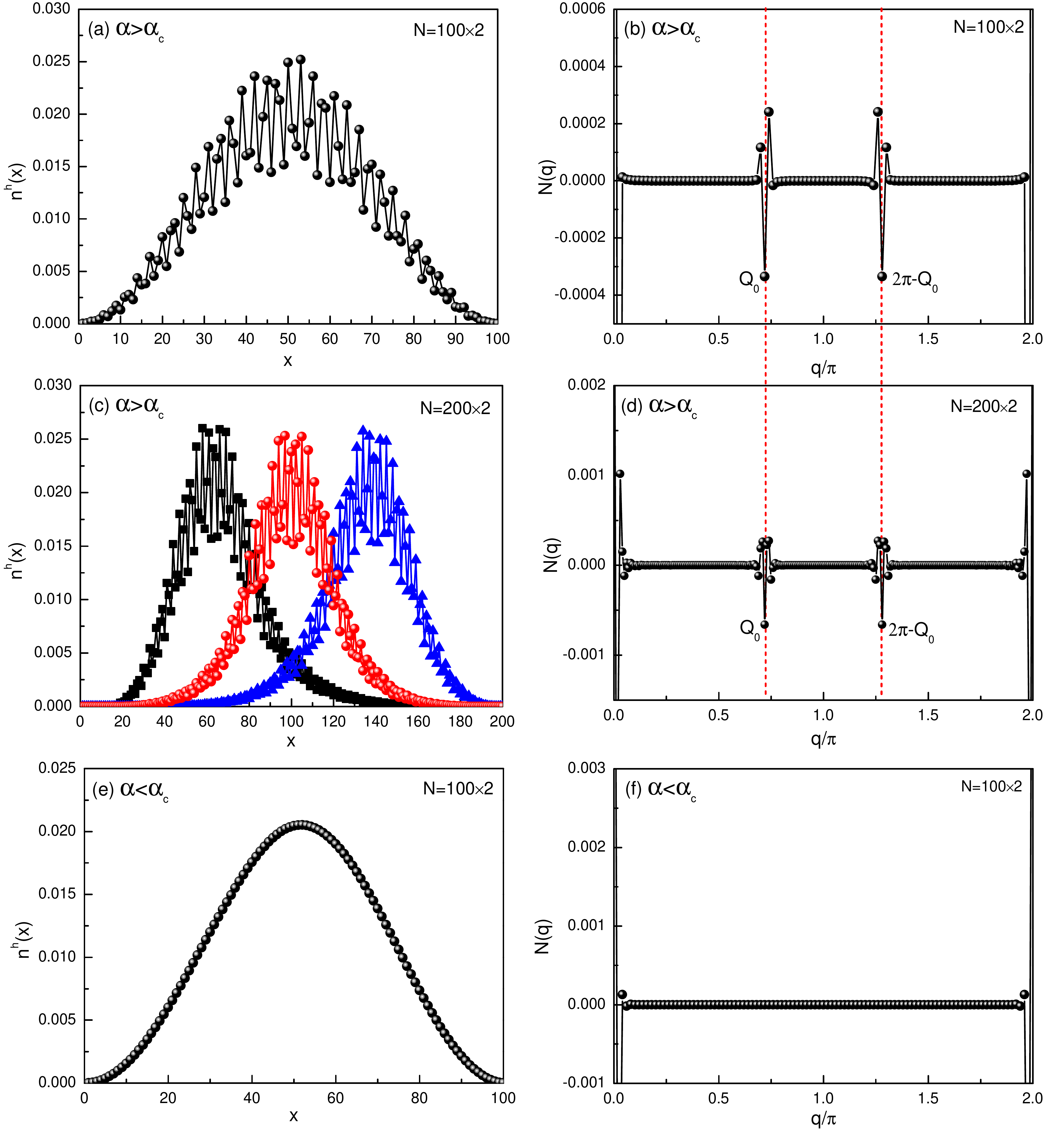}
    }
\caption{(Color online)  (a) Simulation results show that for $\alpha>\alpha_c$ the hole density profile across a two-leg $t-J$ ladder exhibits a charge modulation, a (spatial) oscillation, which superposes on a smooth background localized at the sample center. (b) The Fourier transformation of the profile exhibits two peaks at $Q_0$ and $2\pi-Q_0$, implying that the oscillatory pattern has a period of $2\pi/Q_0$. (c) Upon introducing weak impurities the hole density ``collapses'' into a localized profile pinned at the localization center \cite{ZZ2013}.
(d) The profile shifted with the localization center notwithstanding, the peaks of the Fourier transformation of the profile remain unchanged, implying the invariance of the period of the charge modulation.
(e) Simulations of the same system but with $\alpha<\alpha_c$ show the absence of charge modulations concomitant with the hole delocalization \cite{ZZ2014qp}. (f) Correspondingly, the Fourier transformation of the hole density profile does not exhibit any peaks. Here, $t/J=3$ and for this ratio it has been found \cite{ZZ2014qp} that $\alpha_c=0.7$. For (a)-(d) $\alpha=1$ while for (e) and (f) $\alpha=0.4$.}
\label{Fig:HoleDensity}
\end{figure*}

In essence, at small length scales the phase strings give rise to a coherent propagation of the hole through the quantum spin background
characterized by sharp single-particle momenta, with the momentum value set by the many-body physics of the quantum spin background.
This coherent propagation is however confined on larger scales by the localization volume that acts like a box. This has in turn the consequence
that the interference between the left- and right-propagating modes of the hole inside this volume can form standing
waves corresponding to the charge modulations.

\begin{figure}[tbp]
\centerline{\includegraphics[height=2in,width=2.8in] {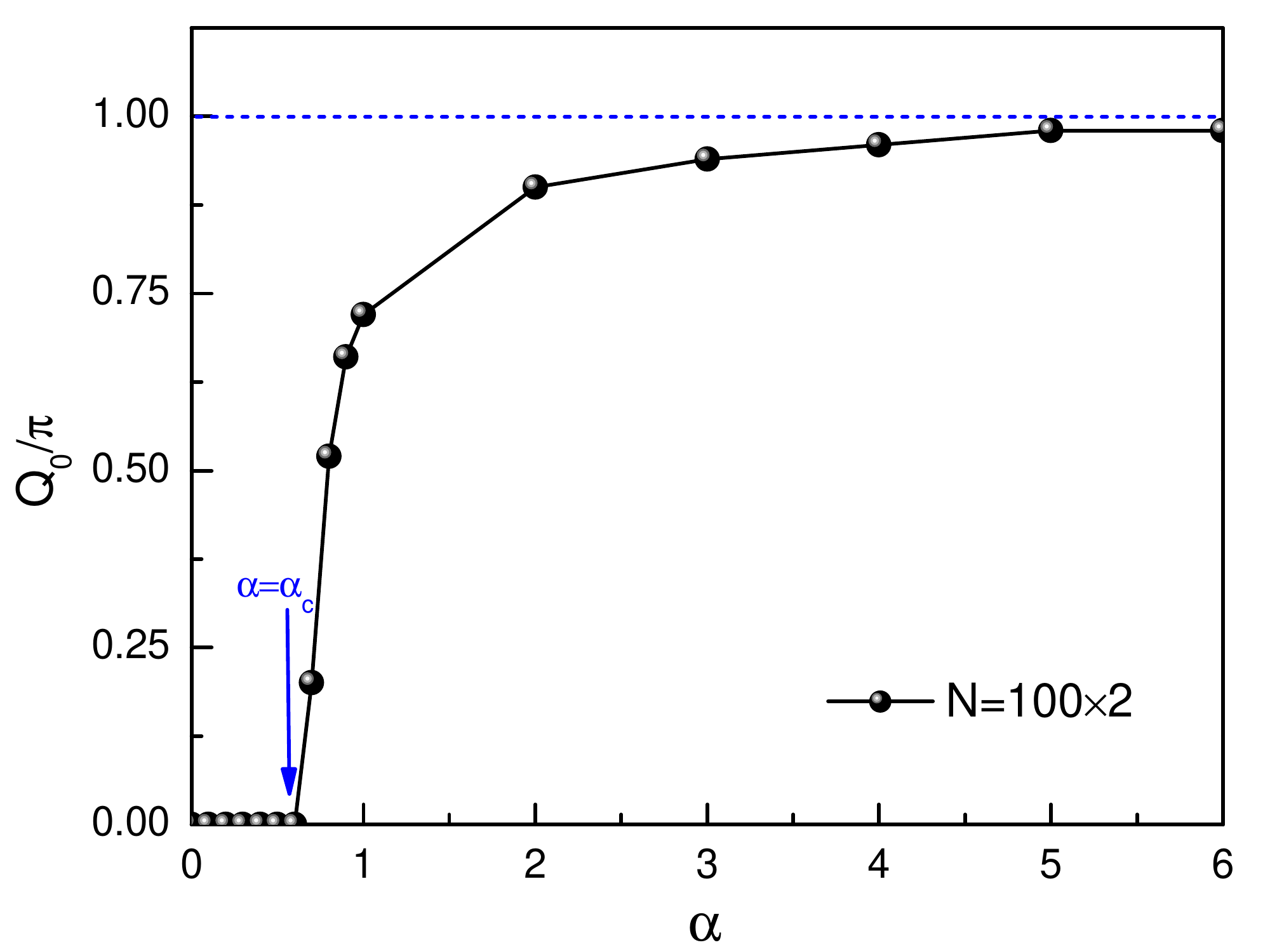}}
\caption{(Color online) Simulation results of $Q_0$ for different values of the rung coupling $\alpha$. $Q_0$ is found to vanish for $\alpha<\alpha_c=0.7$. Above the threshold it increases monotonically with $\alpha$ and is found to approach $\pi$ in the large $\alpha$ limit.}
\label{Fig:Q0}
\end{figure}

To convince the reader we  will start out presenting numerical DMRG results in section \ref{sec:t-J_model}
 for a two-leg $t-J$ ladder system doped with a single hole (Fig.~\ref{Fig:Model}) after introducing the model and the numerical method in section II.
To unambiguously identify the origin of the spontaneous charge modulations found in the numerical results we will employ the same strategy as used in the previous work
\cite{ZZ2013}:
By adding a simple spin dependence to the hopping term [see Eq.~(\ref{eq:7})] the effects of the phase strings are canceled out.
Comparing the outcomes with those of the standard $t-J$ model one immediately identifies the modulation of the hole density [Figs.~\ref{Fig:HoleDensity}(a)-(d)] as being caused by the phase strings.  The highly unconventional origin of the charge modulation is also unveiled by
its dependence on the strength of the rung coupling $\alpha$ (cf. Fig.~\ref{Fig:Model}). This effectively tunes the density of spin flips and thereby the density
of the phase string signs. For small $\alpha < \alpha_c$ the hole delocalizes and the modulation disappears [Figs.~\ref{Fig:HoleDensity}(e)
and \ref{Fig:HoleDensity}(f)]. It is found that in  the self-localized regime
($\alpha > \alpha_c$) the modulation period rapidly increases with $\alpha$ (Fig.~\ref{Fig:Q0}).
In section~\ref{sec:interference} we will rest on the analogy with the theory
of Anderson localization, devising a path-integral method that makes it possible to analyze the interference phenomena caused by the phase strings
in an analytical fashion. This revolves around the assertion that it is possible to assign an {\em average} phase string sign to any hopping path of the
hole, { essentially a mean-field like approximation. This} maps the problem of the hole propagation on an effective quantum mechanics problem which is amenable to a path integral treatment, be it that
the interference phenomena are now rooted in these phase string signs.

From this construction we identify an analog of the mechanism for the familiar Anderson localization, to subsequently unveil the origin of the spontaneous charge
modulation in terms of interference of emergent left- and right-movers inside the localization volume. This analysis reveals the cause of the highly surprising
dependence of the modulation  period  on the rung coupling: It is set by the probability $p_{\uparrow}$ for the hole (injected by removing a $\downarrow$-spin electron) to exchange its position with
an up($\uparrow$)-spin nearest neighbor, a quantity that measures effectively the density of phase string signs in the spin vacuum.
In section~\ref{sec:discussion} we will end with an outlook, discussing the potential ramifications of this mechanism for charge modulation in the broader
context of the physics of doped Mott insulators. Some technical details are relegated to Appendix~\ref{sec:band}.

\section{Models and numerical methods}
\label{sec:model_and_method}

Let us depart from the standard $t-J$ model describing the basic physics of a doped Mott-insulator in the strong coupling limit, with
Hamiltonian  $ H_t+H_J$, where $H_t$ describes that the hole hops from one site to the other and $H_J$ is the superexchange interaction between nearest spins. These two terms read
\begin{eqnarray}
    H_t &=& -\sum_{\langle ij\rangle\sigma}t_{ij}\left(c_{i\sigma}^\dagger c_{j\sigma}+h.c.\right), \label{eq:3}\\
    H_J &=& \sum_{\langle ij\rangle}J_{ij}\left({\bf S}_i\cdot {\bf S}_j-\frac{1}{4}n_in_j\right),
    \label{eq:4}
\end{eqnarray}
respectively. Here $t_{ij}>0$ is the hopping integral and $J_{ij}>0$ the superexchange coupling. $c_{i\sigma}$ is the electron annihilation operator at site $i\equiv (x,y)$, with $\sigma=\uparrow, \downarrow$ being the spin index, ${\bf S}_i$ the spin operator,
$n_i=\sum_{\sigma}c_{i\sigma}^\dagger c_{i\sigma}$ the electron number operator, and $\langle ij\rangle$ denotes the nearest neighbors.
The Hilbert space is constrained by the no-double occupancy condition, i.e., $n_i \leq 1$. We consider the system doped by a single hole, i.e.,
$\sum_i n_i=N-1$ where $N$ is the number of lattice sites. In our numerical simulations, the single-hole doping is realized by removing a $\downarrow$-spin electron.

We focus on the ladder geometry of length $N_x$ and leg number $N_y=2$ (Fig.~\ref{Fig:Model}): This is the so-called rung model.
 For simplicity we consider even $N_x$ throughout this work.  Such  $t-J$ ladders have been recently employed with much success to establish
 numerically the phase string-triggered self-localization effect \cite{ZZ2013} and the associated collapsing of the quasiparticle \cite{ZZ2014qp}.
These ladders are  particularly convenient for numerical investigations using the DMRG method \cite{White92}.
The interchain hopping and superexchange parameters are $t$ and $J$, respectively, while their intrachain counterparts are $\alpha t$ and $\alpha J$, $\alpha>0$. In this case, Eqs.~(\ref{eq:3}) and (\ref{eq:4}) are simplified to
\begin{equation}
H_{t} =  - t\sum_{x,y,\sigma } \left(c_{x,y,\sigma }^\dag c_{x,y+1,\sigma }
+\alpha c_{x,y,\sigma }^\dag {c_{x + 1,y,\sigma }}\right) + {\rm h.c.},
\label{eq:5}
\end{equation}
with the abbreviation `h.c.' being the Hermitian conjugate, and
\begin{eqnarray}
H_{J} &=&  J\sum_{x,y,\sigma } \bigg(({\bf S}_{x,y}\cdot {\bf S}_{x,y+1}-\frac{1}{4}n_{x,y}n_{x,y+1})\nonumber\\
&&\quad +\alpha ({\bf S}_{x,y}\cdot {\bf S}_{x+1,y}-\frac{1}{4}n_{x,y}n_{x+1,y}) \bigg),
\label{eq:6}
\end{eqnarray}
where $x$ denotes the coordinate along the chain direction and $y$ the leg index. { (We note that for the more physical Hubbard model with $U/t\gg 1$, one should replace the anisotropic parameter $\alpha$ in $H_J$ by $\alpha^2$. These two models seem to differ greatly in the limit of $\alpha\gg 1$. But our preliminary investigations have shown that the results are qualitatively the same as those to be presented below provided that $\alpha$ is not much larger than one.)}

The weapon of choice to isolate the effects of the phase strings is associated with the following modification of the hopping term, i.e.,
replacing  Eq.~(\ref{eq:5})  by
\begin{equation}
H'_{t} =  - t\sum_{x,y,\sigma } \sigma\left(c_{x,y,\sigma }^\dag c_{x,y+1,\sigma }
+\alpha c_{x,y,\sigma }^\dag {c_{x + 1,y,\sigma }}\right) + {\rm h.c.}.
\label{eq:7}
\end{equation}
As discussed in Ref.~\cite{ZZ2013}, the effect of making the sign of the hopping spin($\sigma$)-dependent is precisely to cancel out the phase strings,
and by comparing with the outcomes of the standard $t-J$ model one can isolate the unique effects of the phase strings in an unambiguous way
 (see section~\ref{sec:interference} for a detailed analysis).

Along these lines, it was found in Ref. \cite{ZZ2013} that the self-localization of an injected hole occurs in $t-J$ ladders but not in $\sigma\cdot t-J$ ladders.
More precisely, it has been shown very recently \cite{ZZ2014qp} that for the $t-J$ ladder there is a critical value $\alpha_c$ above which the injected hole is
self-localized, with the consequence that the quasiparticle picture of the hole is invalidated.
In sharp contrast, for the $\sigma\cdot t-J$ model the hole doped into the ladder behaves invariably in a quasiparticle manner for all values of $\alpha$.

Here we focus on another surprising feature that was overlooked in the earlier works \cite{ZZ2013,ZZ2014qp}:
the presence of spatial modulations of the hole density in the DMRG results
for the standard $t-J$ ladders.  One could be tempted to associate these with Friedel oscillations coming from the open boundaries. However, the comparison
with the $\sigma\cdot t-J$  results leaves no room for ambiguity: These modulations completely disappear when the phase strings are removed.
Moreover, even for standard $t-J$ ladders we find that these charge modulations only occur when $\alpha>\alpha_c$, i.e., in the self-localized regime.

We use in the remainder  of this paper the standard DMRG method to simulate the ground state of these two systems. In doing so,
we keep $300 \sim 5000$ states in the DMRG block with around $10\sim40$ sweeps to obtain converging results. The truncation error is $\lesssim 10^{-8}$.
\begin{figure}[tbp]
\centerline{
    \includegraphics[height=3.8in,width=2.8in] {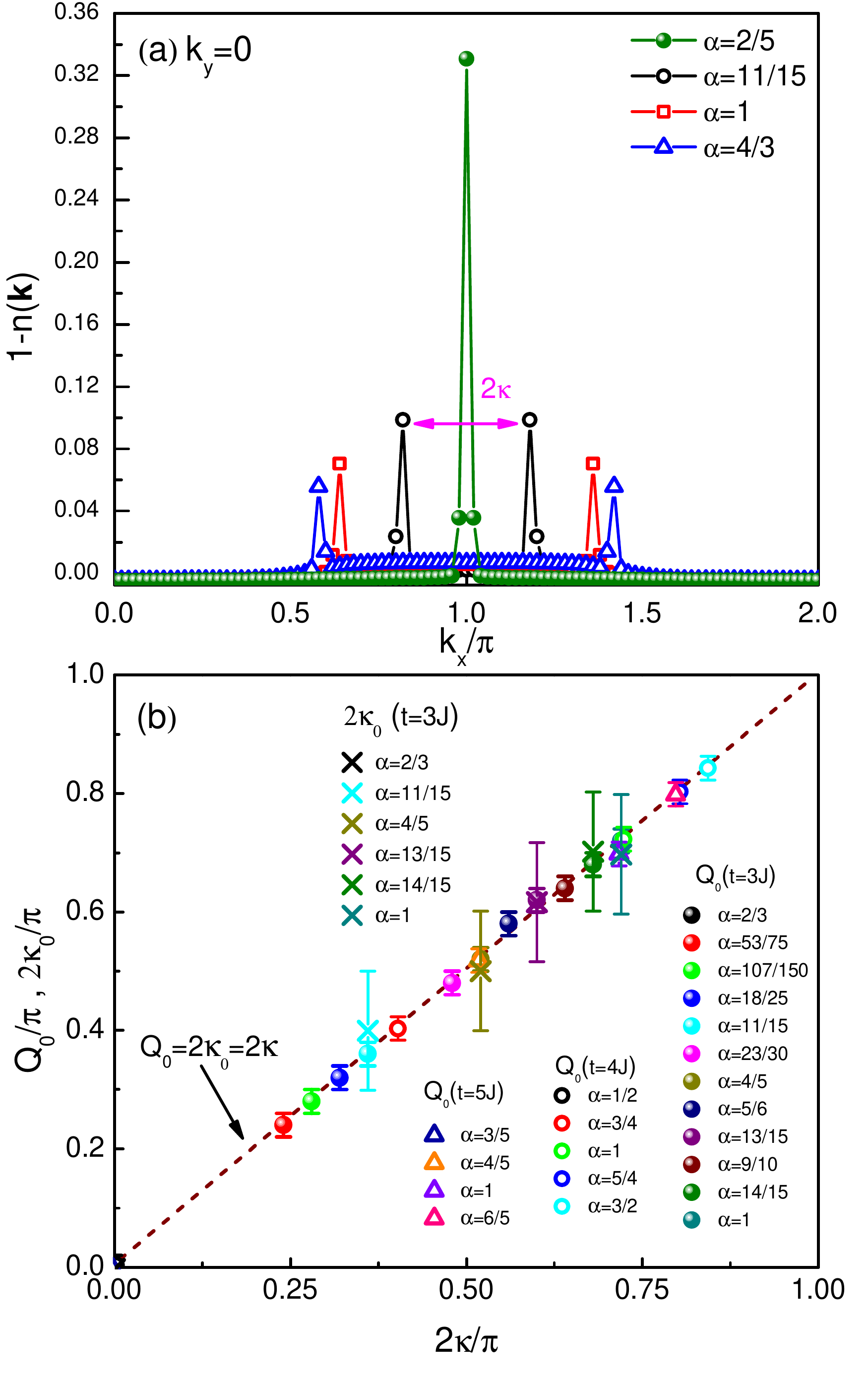}
    }
\caption{(Color online) (a) The single-hole momentum distribution ($k_y$=0) for different values of $\alpha$ in the two-leg $t-J$ ladder, where $t/J=3$ and the critical value $\alpha_c=0.7$. For $\alpha$ below $\alpha_c$ the distribution displays a quasiparticle peak at $k_x=\pi$; above $\alpha_c$ this quasiparticle peak splits into two symmetric with respect and of a distance $\kappa$ to $k_x=\pi$. (b) The wave number of the charge modulation $Q_0$ and the peak distance of the single-hole momentum distribution $2\kappa$ are found to be identical to twice the wave number of the oscillation of $\Delta E_G^{\rm 1-hole}(N_x)$, i.e., $Q_0=2\kappa=2\kappa_0$ (dashed line). Note that $Q_0=2\kappa=2\kappa_0\equiv 0$ at $\alpha<\alpha_c$.
}
\label{Fig:Scaling}
\end{figure}
\begin{figure}[tbp]
\centerline{
    \includegraphics[height=3.8in,width=2.8in] {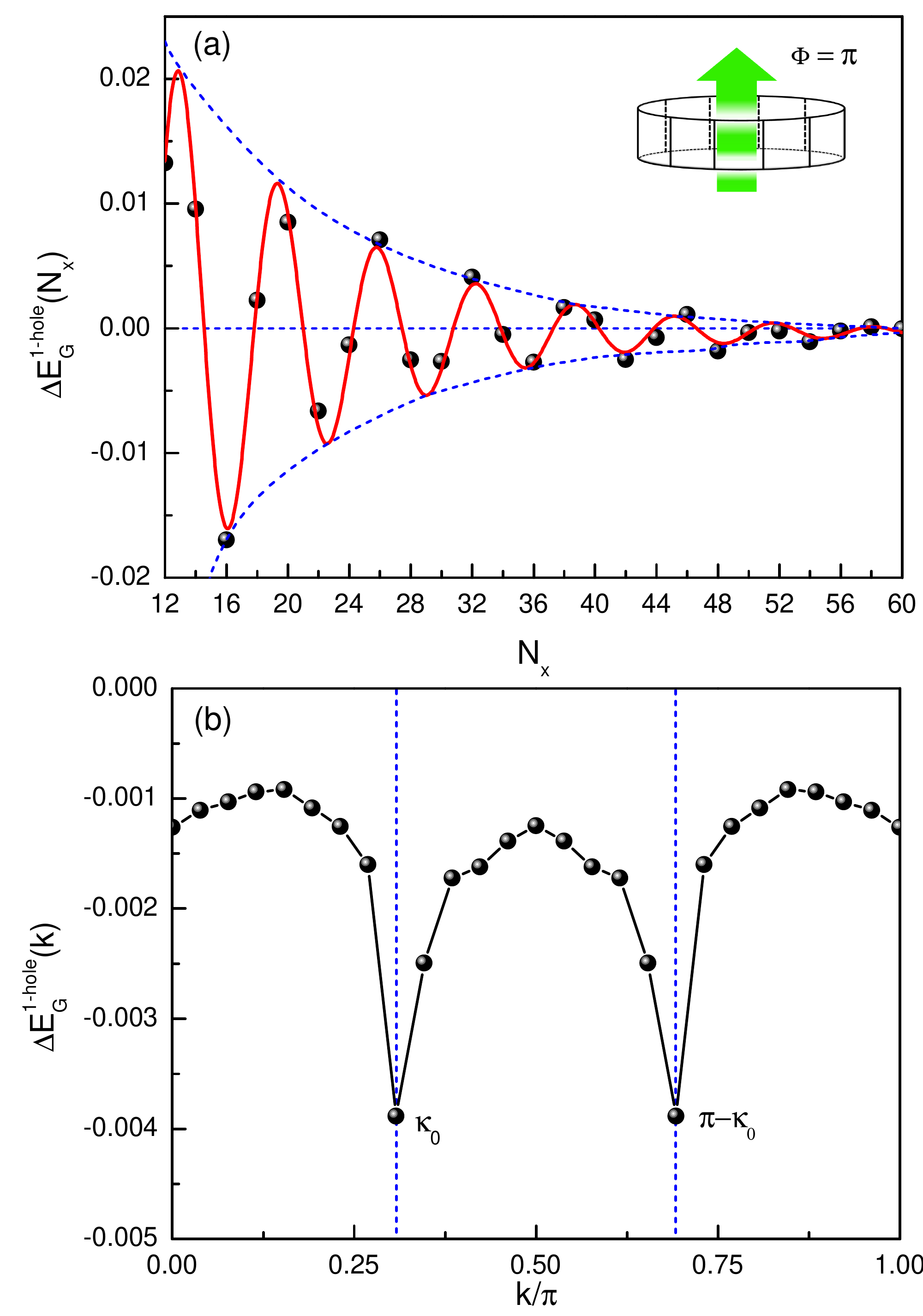}
    }
\caption{(Color online) (a) The energy difference $\Delta E_{G}^{\text{1-hole}}(N_x)$ of a single-hole doped two-leg $t-J$ ladder displays an oscillation in $N_x$, with the amplitude decaying exponentially when $N_x$ approximately is larger than $12$. To calculate the energy we compactify the ladder by ``gluing" its two ends and let a magnetic flux of $\Phi=\pi$ pierce through it (upper right). (b) The Fourier transformation of this energy difference exhibits two peaks at $\kappa_0$ and $\pi-\kappa_0$, respectively. Here, $\alpha=13/15$ and $t/J=3$.}
\label{Fig:Edifference}
\end{figure}

\section{Numerical results}
\label{sec:t-J_model}

The hole density profile across the ladder is defined as
\begin{equation}
n^h(x) \equiv 1- \sum_{y,\sigma} {\langle c^\dagger_{x,y,\sigma}c_{x,y,\sigma}\rangle},
\label{eq:8}
\end{equation}
where $\langle\cdot\rangle$ denotes the ground-state average. The ground state is computed  using the DMRG method implemented with
open boundary conditions, from which the hole density  is computed.

\subsection{The $t-J$ ladders}
\label{sec:numerics_t-J_ladder}

\subsubsection{The self-localized regime}
\label{sec:localization}

For  $\alpha > \alpha_c$ the hole is known to be self-localized from earlier numerical studies \cite{ZZ2013,ZZ2014qp}, as mentioned in the above.
We present in Fig. \ref{Fig:HoleDensity}(a)
a typical result for the hole density profile along the ladder. One directly infers the presence of an oscillatory pattern which is superposed on a smooth profile.
We computed  the Fourier transformation of this profile, defined as
\begin{equation}
N(q) \equiv \frac{1}{N_x}\sum\limits_{x=0}^{N_x-1} n^h(x) {e^{iqx}},
\label{eq:9}
\end{equation}
where the Fourier wave number $q=2\pi m/{N_x},m=0,1, \cdots , N_x-1$. We find that $N(q)$ exhibits two sharp peaks.
Their positions are denoted by $Q_0$ and $2\pi-Q_0$, respectively [cf. Fig.~\ref{Fig:HoleDensity}(b)]. It follows that the charge density
corresponds to a single harmonic modulation and therefore must oscillat periodically (in space) with a period of $2\pi/Q_0$. Our simulation results further show that $Q_0$ increases
monotonically,  converging eventually  to the limit $Q_0=\pi$ for $\alpha\rightarrow\infty$ (Fig.~\ref{Fig:Q0}).

Interestingly, impurities shuffle the modulation pattern, but leave its wave number/period unaffected [cf. Figs.~\ref{Fig:HoleDensity}(c) and \ref{Fig:HoleDensity}(d)]. This indicates that the charge modulation is intrinsic to the hole self-localization, rather insensitive to  the environment
outside the localization volume, and occurs essentially inside this volume.
Notice that the localization profile in Fig.~\ref{Fig:HoleDensity}(c) is narrower than the profile in Fig.~\ref{Fig:HoleDensity}(a). This is because the latter is the incoherent superposition of localized profiles characterized by different localization centers \cite{ZZ2013}.

To further investigate the relation between the charge modulation and the self-localization we compute the single-hole
momentum distribution, defined as $1-\sum_{\sigma } \langle {c_{\mathbf{k}\sigma}^{\dag }c_{\mathbf{k}\sigma }}\rangle \equiv 1-n(\mathbf{k})$,
where $c_{\mathbf{k}\sigma}^{\dag }$ and $c_{\mathbf{k}\sigma}$ are the Fourier transforms of $c_{i\sigma}^{\dag }$ and $c_{i\sigma}$, respectively.
As is seen in Fig.~\ref{Fig:Scaling}(a),  the quasiparticle picture collapses for $\alpha>\alpha_c$: The quasiparticle peak centered at $k_x=\pi$ disappears, split into two small peaks at $k_x=\pi \pm \kappa$ which appear symmetrically relative to
$k_x=\pi$. We stress that these two peaks are not quasiparticle peaks because
 they vanish in the thermodynamic limit $N_x\rightarrow \infty$. They represent instead left- and right-moving hole waves with momentum of ($\pi \pm \kappa$) inside
the localization volume. Our simulations reveal in addition that the distance $2\kappa$ between these two peaks coincides with the wave number of
the charge modulation $Q_0$ [Fig.~\ref{Fig:Scaling}(b)]. This suggests that the charge modulation is caused by the interference between
the left- and right-moving waves. This will be further elucidated in section~\ref{sec:interference} using analytic methods.

Finally, we also studied the influence of periodic and antiperiodic boundary conditions on the
ground-state energies. We define the difference in energy caused by these two different boundary conditions as $\Delta E_G^{\rm 1-hole}(N_x)$
and results  are shown in Fig.~\ref{Fig:Edifference}(a). The envelope of $\Delta E_G^{\rm 1-hole}(N_x)$ turns out to
 decay exponentially with $N_x$, consistent with earlier findings \cite{ZZ2013}. More importantly, this energy difference
 displays an oscillation in $N_x$ including a single harmonic: as shown in Fig.~\ref{Fig:Edifference}(b)
 its Fourier transformation, denoted as  $\Delta E_G^{\rm 1-hole}(k)$, exhibits two peaks at $\kappa_0$ and $\pi-\kappa_0$.
 As inferred from Fig.~\ref{Fig:Scaling}(b), $\kappa_0=\kappa$.

\subsubsection{The delocalized regime}
\label{sec:delocalization}

It is known that for $\alpha<\alpha_c$ the hole is delocalized, behaving like a quasiparticle \cite{ZZ2014qp}, while it has a
momentum $k_x=\pi$  as revealed by  the central peak of Fig.~\ref{Fig:Scaling}(a).
According to  Fig.~\ref{Fig:HoleDensity}(e) (real space) and   Fig.~\ref{Fig:HoleDensity}(f) (momentum space) the charge modulations disappear
completely in this regime.

\subsection{The $\sigma\cdot t-J$ model}
\label{sec:sigma_t-J_model}

\begin{figure}[tbp]
\centerline{   \includegraphics[height=5.6in,width=3.0in] {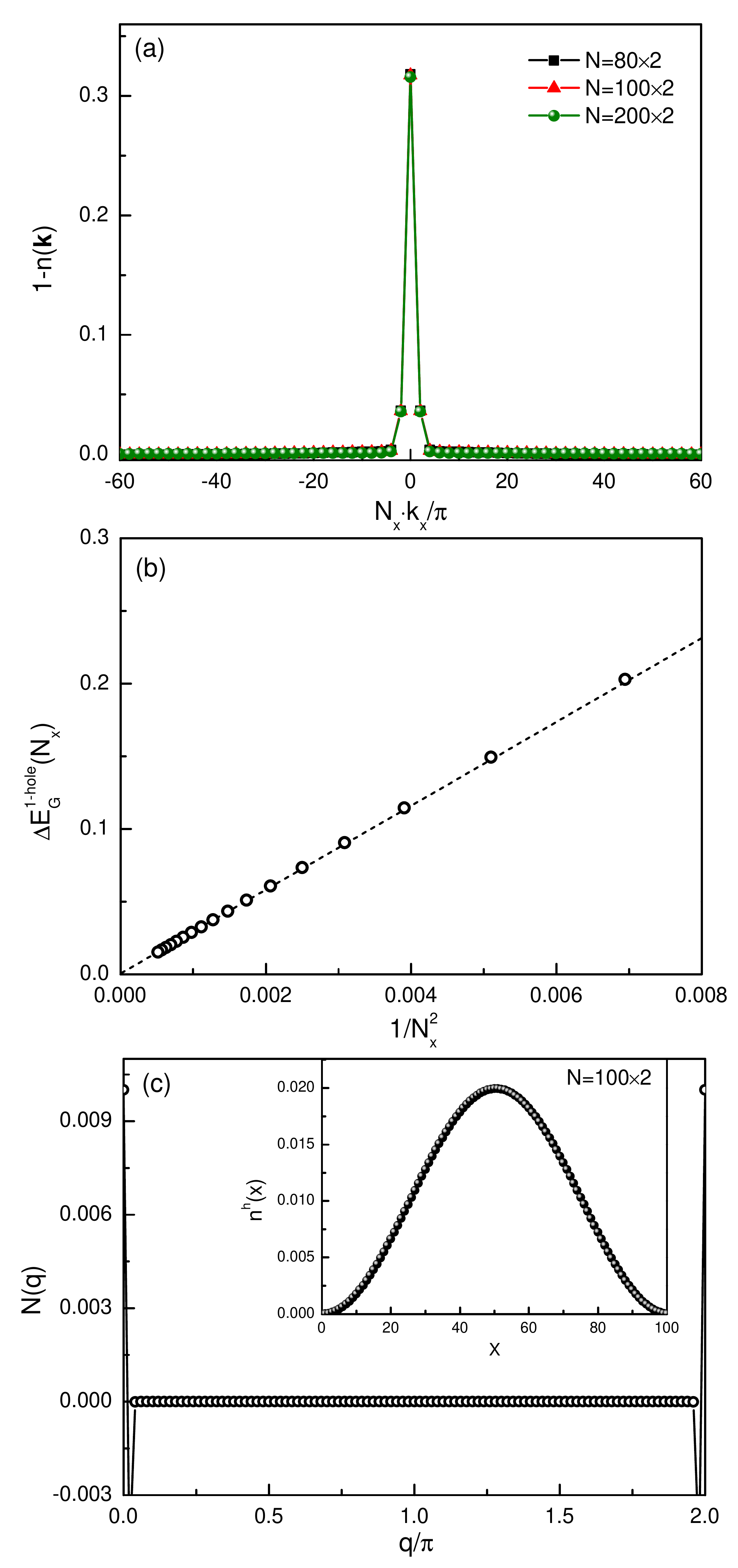}}
\caption{(Color online) The simulation results for the single-hole doped $\sigma\cdot t-J$ ladder. (a) The single-hole momentum distributions for different ladders fall into a universal curve displaying a sharp quasiparticle peak. (b) The hole's energy also displays a ``free-particle" like behavior, i.e., $\Delta E_G^{\rm 1-hole}\sim N_x^{-2}$. (c) Both the spatial resolution (inset) and the Fourier transformation (main panel) of the hole density profile show the absence of charge modulations. Here, $\alpha=1$ and $t/J=3$.} \label{Fig:sigma_t-J}
\end{figure}

To prove that both the self localization and the charge modulations are due to the phase strings we repeated the computations for the
single-hole doped $\sigma\cdot t-J$ ladders. The  main differences between the $t-J$ and $\sigma\cdot t-J$ ladders are summarized in Table~\ref{tab:1}.
Consistent with the previous work \cite{ZZ2013}, these results demonstrate that in the absence of the phase string signs the injected hole behaves
always as an  impeccable quasiparticle with its spectrum exhibiting a sharp quasiparticle peak,  irrespective of $\alpha$  and the number of rungs $N_y$.
In fact, we find that upon rescaling according to  $k_x \rightarrow N_x k_x$,
the single-hole momentum distribution for different ladders collapses onto a universal curve  [Fig.~\ref{Fig:sigma_t-J}(a)]  as long as the ladders  are sufficiently long,
while the finite-size scaling of the  energy of the hole reflects that it propagates freely,
 $\Delta E_G^{\rm 1-hole}\sim N_x^{-2}$ [Fig.~\ref{Fig:sigma_t-J}(b)]. What matters most in the present context is there is  no  sign of charge modulations
 in the absence of the phase strings as illustrated in Fig. \ref{Fig:sigma_t-J}(c). Let us now turn to the path integral considerations, which reveals the physics
 principle explaining these puzzling observations.

\begin{table}
\newcommand{\tabincell}[2]
{\begin{tabular}{@{}#1@{}}#2
\end{tabular}}
\centering
\caption{\label{tab:1}Comparison of main properties of two kinds of single-hole doped two-leg ladders. The notation $+$ ($-$) stands for the presence (absence) of phenomena listed in the first column.}
\begin{tabular}{c|c|c|c}
\hline
\hline
\multirow{2}{*}{} & \multicolumn{2}{c|}{$t-J$} &
$\sigma\cdot t-J$\\
\cline{2-4}
&$\alpha<\alpha_c$ & $\alpha>\alpha_c$ & $\alpha>0$ \\
\hline
quasi-particle & $+$ & $-$ & $+$ \\
\hline
self-localization & $-$ & $+$ & $-$ \\
\hline
charge modulation & $-$ & $+$ & $-$ \\
\hline
phase strings & $+$ (but canceled out) & $+$ & $-$ \\
\hline
\hline
\end{tabular}
\label{tab:1}
\end{table}

\subsection{The spin response}
\label{sec:spin}

\begin{figure}[tbp]
\centerline{   \includegraphics[height=3.8in,width=2.8in] {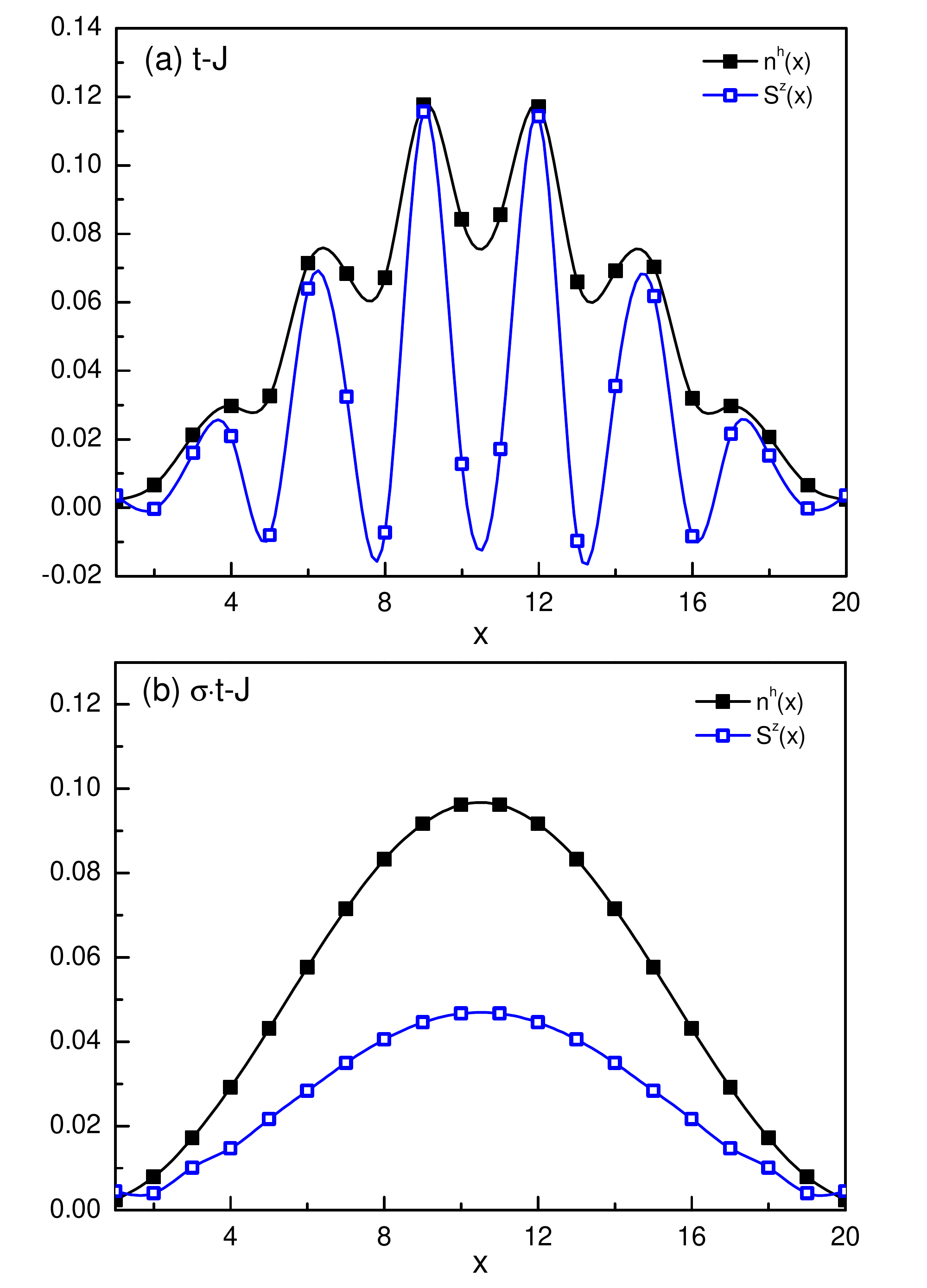}}
\caption{(Color online) The simulation shows that in $t-J$ ladders the charge modulation is accompanied by the spin modulation (a) while in $\sigma\cdot t-J$ ladders they both disappears. Here, $\alpha=1$ and $t/J=3$.}
\label{Fig:spin_modulation}
\end{figure}

{ So far we have only considered the charge response to the single-hole injection. Note that the injection also creates an $\uparrow$-spin (by removing a $\downarrow$-spin). Because of the coupling between charge and spin degrees of freedom, we expect that the charge modulation could have prominent consequences on the spin degree of freedom. Specifically, we also simulate the spin density,
\begin{equation}
S^z(x) \equiv \sum_{y,\sigma} {\sigma\langle c^\dagger_{x,y,\sigma}c_{x,y,\sigma}\rangle}.
\label{eq:15}
\end{equation}
As shown in Fig.~\ref{Fig:spin_modulation}, in the $t-J$ ladders the charge modulation generally leads to a spin modulation. Consistent with this, in the $\sigma\cdot t-J$ ladders where the charge modulation disappears, the spin modulation is not observed.

We remark that a modulated spin cloud has been known to accompany a static hole localized on a single site in a two-leg ladder and such modulations have been observed in experiments on Zn
doped cuprate ladders (for a review see, e.g., Ref.~\cite{Dagotto99}). This corresponds to the elimination of the hole's hopping in the $t-J$ and $\sigma\cdot t-J$ models. As such, these two models are reduced to the same Heisenberg spin ladders with an empty site, and a spin-$1/2$ will distribute around the impurity site with a short-ranged antiferromagnetic oscillation due to the underlying spin-spin correlation. Once the hole's hopping is restored, the modulated spin cloud is carried by the hole, which is smoothened in the $\sigma \cdot t-J$ ladder since the hole's distribution is smooth. By contrast, in the $t-J$ ladder, because the hole's density acquires a spatial modulation in the chain direction, the accompanying spin-$1/2$, with its dominant amplitude at the opposite site of the same rung in the static limit, exhibits a similar modulation. In other words, the spin modulation found here is directly related to the charge modulation due to the spin-charge correlation. This also explains why the wavenumbers of these two modulations are approximately the same, in contrast to many conventional systems in which the charge modulation is usually observed to be
the second harmonic of a spin modulation. }

\section{Phase-string world histories: charge modulation and wave interference}
\label{sec:interference}

In the exposition of the numerical results it shimmers through that somehow the hole injected into the $t-J$ model yields to the principles of free-particle
quantum mechanics. However, very different from the ``simple"  wave interference mechanisms behind  Anderson localization and the Friedel
oscillations we are now dealing with a strongly interacting system. The  very origin of the interference phenomena encountered in this
context is rooted in many-particle physics of a particularly hairy kind -- it is about the way that fermion signs do their work in an extremely strongly
coupled fermionic system. It has already been recognized for a long while that these can be conveniently enumerated in terms of the phase strings in
the case of the $t-J$ model  \cite{Sheng1996,Weng1997,weng_07}, but this does not solve the problem in general terms. In this section we will demonstrate
that for the particular problem of a single hole we can keep track of the way that these phase string signs accumulate in world histories in terms of a
simple mean-field like average associated with long hopping paths of the hole. Departing from this assertion,
the sum over histories  turns into an effective quantum-mechanics like wave interference affair that in turn explains semi-quantitatively
not only the self-localization  but also the highly structured behavior in single hole momentum space that is responsible for
 the charge modulations observed in the numerical DMRG results.  We stress that there is no precedence elsewhere for this physics;
the superficial similarities with simple quantum mechanics is deceptive in this regard.

We depart from the single-hole Green's function, defined as
\begin{eqnarray}\label{eq:16}
    G(j,i;E)
    \equiv \langle \psi_0|c_{j\downarrow}^\dagger (E-H_t-H_J
    )^{-1}c_{i\downarrow}|\psi_0\rangle,
\end{eqnarray}
where $\psi_0$ is the ground state at half filling. Physically, it describes the propagation of an injected hole (realized by removing a $\downarrow$-spin electron)
from site $i$, to site $j$.  According to an exact theorem that applies to the $t-J$ model \cite{Sheng1996,Weng1997,weng_07},
for $E<0$ (which is satisfied in this context) this Green's function can be expressed as
\begin{eqnarray}
G(j,i;E)=(-1)^{j-i+1}\sum_p A_p e^{iS_p}, \quad S_p=\pi N_p^\downarrow,
\label{eq:17}
\end{eqnarray}
where the summation is over all the paths $p$ connecting $i$ and $j$, $N_p^\downarrow$ counts the number of  times
that the hole exchanges its position with $\downarrow$-spins along a given path $p$, while the amplitude $A_p>0$
is determined by all intermediate spin configurations. Among others, this differs fundamentally from the standard first quantized
path integral of quantum mechanics by the following fact:  The ``action'' is associated with the hole acquiring a quantum phase $\pi$,
every time that it exchanges its position with a $\downarrow$-spin. Each world history thereby carries a unique phase $S_p$ that enumerates
the sum of these phases associated with this history, which are the so-called
phase strings \cite{Sheng1996,Weng1997,weng_07}. These correspond to oscillatory factors of a new kind appearing in the path integral, that survive
even in Euclidean signature. We will see that these underlie the ``quasi-quantum mechanical" interference phenomena.

From this formulation it is immediately clear why the phase strings are completely canceled in the $\sigma\cdot t-J$ model, turning it into the powerful  tool
to isolate their specific influences in the numerical simulations (Fig.~\ref{Fig:sigma_t-J}).
The additional sign in the hopping term (\ref{eq:7}) precisely compensates for the quantum phase  $\pi$, generated upon the exchange of the positions of the hole and the $\downarrow$-spin. This just amounts to the statement that $S_p$ in Eq.~(\ref{eq:17}) vanishes identically,
\begin{eqnarray}
S_p=(\pi-\pi) N_p^\downarrow=0.
\label{eq:28}
\end{eqnarray}

\begin{figure}[tbp]
\centerline{\includegraphics[height=3.2in,width=2.8in] {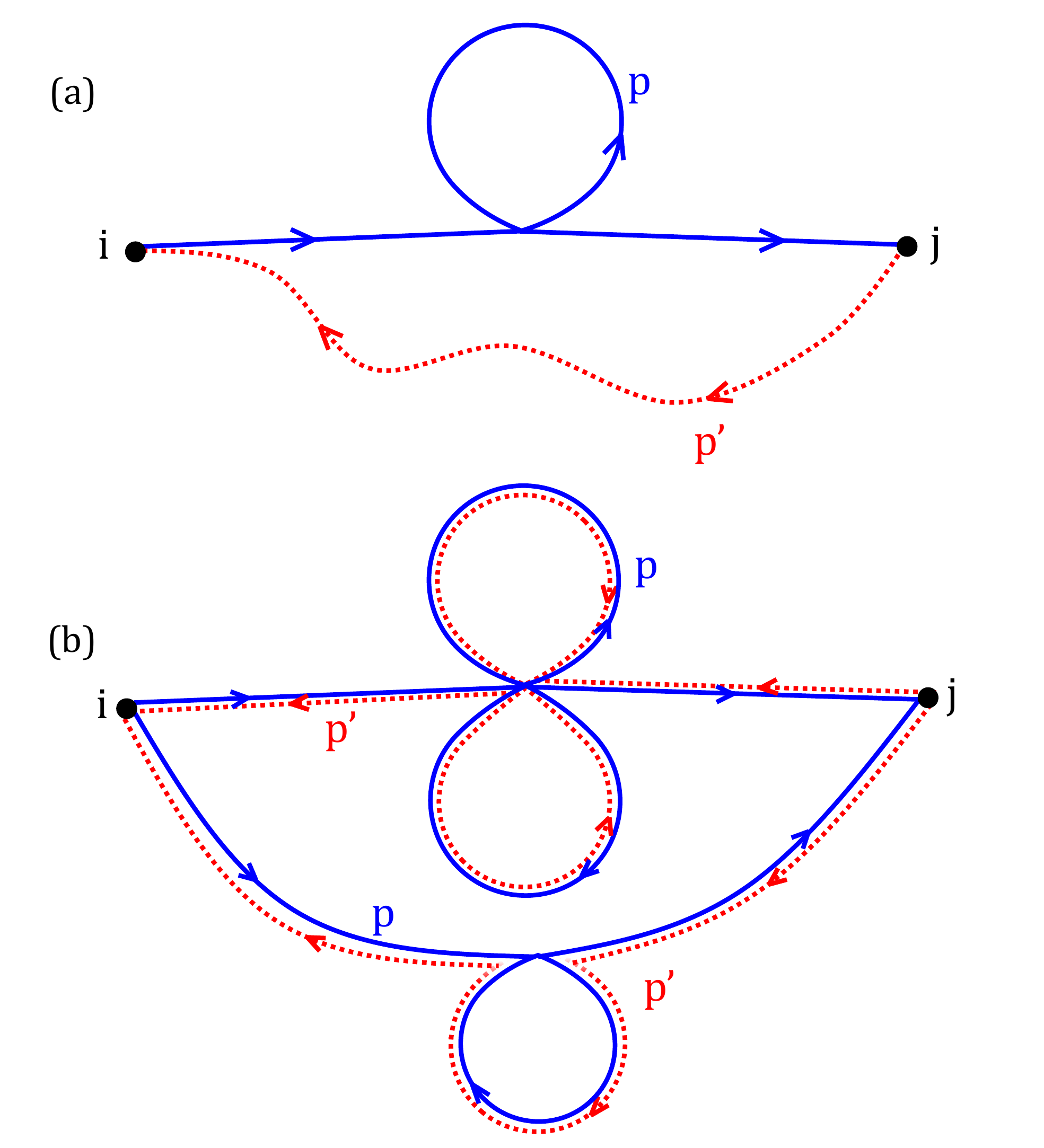}}
\caption{Interference between the paths $p$ (blue solid curve) and $p'$ (red dashed curve), where $p$ ($p'$) represents the ``quantum amplitude'' corresponding to (the complex conjugate of) the single-hole Green's function. (a) For $p_\downarrow=1$ all path pairs $(p,p')$, $p \neq p'$, constructively interfere with each other. This gives rise to the delocalization of the injected hole. (b) Examples lead to constructive interference between path pairs $(p,p')$, $p \neq p'$ for generic values of $p_\downarrow$, giving rise to the self-localization of the injected hole.}
\label{Fig:interference}
\end{figure}

\subsection{The statistical averaging of phase strings and the self-localization
}
\label{sec:self_localization}

A priori one is dealing with a problem which is in principle very complicated: Different from the quantum mechanical case the phase acquired by the propagating hole
is pending the configuration of the $\downarrow$-spins in the spin background, which in turn is subjected to quantum fluctuations on time scales which are typically short
as compared to the time scales associated with the long hopping paths as of relevance to the self-localization. Yet, { by the law of large numbers we could make a conjecture, i.e.,
{\em the phase strings acquire a
mean value} in the case that the path is long.
Although we have no definitive proof of this conjecture, which turns out to be a difficult task, the very existence of this statistical averaging is intuitively reasonable. Indeed, it reflects that at large time scales quantum spin fluctuations are self-averaged. This lays down a foundation of a mean-field like approximation and effectively generates a spatially homogeneous spin background for the large-scale motion of the hole. In the meanwhile, the self-averaging implies that the quantum phase, $S_p$, no longer needs to be multiple $\pi$.}

These mean values determine in turn the properties of the coherent propagation of the hole while the fluctuations around this
mean just lead to random contributions that
cancel out.  As we will see shortly, such statistical averaging gives rise to a self-localization mechanism which is a
firm analog of the Anderson localization mechanism \cite{Lee85}. Strikingly, in this analog the phase-string mean value plays the role
of the phase accumulated when the electron is multiply scattered by quenched disordered potentials. Even more significantly, we will demonstrate underneath that
 this particular averaging of the ``dynamical signs" appears
to be a necessary condition for the sharp quantization in the single-hole momentum space which underlies the charge modulations.  The line of the argument is as
follows: We will first hypothesize that the phase string averages exist, to subsequently demonstrate that these explain both the self-localization,
the sharp quantization in the single-hole momentum space, and the charge modulations on a semi-quantitative level.

Although we have no definitive proof,  the very existence of phase string averages is intuitively reasonable. The phase strings are determined by the number of
times the hole encounters a  $\downarrow$-spin on a particular hopping path. The crucial ingredient required for the averaging is that in a given spin system
the {\em probability} $p_\downarrow$ for a hole to exchange its position with a $\downarrow$-spin nearest neighbor has a well-defined average value. As we
will discuss underneath, to establish this probability requires considerations which are specific for a particular spin system but this involves typically local
physics that is tractable (e.g., the origin of $\alpha_c$). Armed with this hypothesis of statistical averaging we assert that for a hopping path much longer than the
lattice constant the phase string associated with that path
acquires a mean value, simply by multiplying $p_\downarrow$ with the length of the hopping path,
\begin{eqnarray}
S_p \approx \pi p_\downarrow \int_p ds,\label{eq:18}
\end{eqnarray}
where $ds$ is the line element of the path $p$ connecting $i$ and $j$: The phase associated with the path is just proportional to its length,
with a constant of proportionality $p_\downarrow$ which represents the average of the number of times of the ``minus sign events" along this path.
 Substituting Eq.~(\ref{eq:18}) into Eq.~(\ref{eq:17}) gives
\begin{eqnarray}
G(j,i;E)\sim\sum_p A_p e^{i\pi p_\downarrow \int_p ds}.
\label{eq:20}
\end{eqnarray}
Similar to the case of the canonical path integral of quantum mechanics, the exponent with the complex argument causes the world histories to interfere with each other.
Notice that the $A_p$-amplitudes are according to the theorem (\ref{eq:17}) positive-definite quantities for negative energies and therefore all
low-energy interference phenomena are entirely due to the phase string factors. We will see that this simple recipe suffices to explain all the phenomena that were
revealed by the DMRG simulations.

The ``path-integral" formalism (\ref{eq:20}) makes it { easy to mobilize the well-known interference picture} for the study of Anderson localization \cite{Lee85}. Consider the spatial correlation of the hole density between $i$ and $j$, given by
\begin{eqnarray}
|G(j,i;E)|^2\sim\sum_{p,p'} A_pA_{p'} e^{i\pi p_\downarrow \left(\int_p ds-\int_{p'} ds\right)}\nonumber\\
=\sum_{p} A_p^2 + \sum_{p\neq p'} A_pA_{p'} e^{i\pi p_\downarrow \left(\int_p ds-\int_{p'} ds\right)},\quad
\label{eq:22}
\end{eqnarray}
where the first and second sum  in the second line are called  the diagonal and off-diagonal  contributions, respectively.  The latter describes
the interference between the path pair $(p,p')$, $p\neq p'$, where $p$ ($p'$) represents the ``quantum amplitude'' corresponding to
(the complex conjugate of) the Green's function. Depending on both the value of $p_\downarrow$ and the path pair $(p,p')$ the interference may be
either constructive or destructive.

Before we demonstrate the specific mechanism behind the charge modulations, we can already discern the mechanism of the self-localization from
this path integral formulation:

\begin{itemize}
  \item In the special case of $p_\downarrow=1$, all paths have the same phase (modulo $2\pi$) since they differ in an even number of hoppings.
 Consequently, all path pairs regardless of whether they are diagonal or off-diagonal constructively interfere with each other [Fig.~\ref{Fig:interference}(a)].
 This results in a free motion of the hole: All amplitudes are
 positive in the ground state corresponding with a perfectly delocalized state. This is consistent with the numerical finding of the existence of a quasiparticle
 for $\alpha<\alpha_c$ in the two-leg $t-J$ ladder. In this regime  the hole is bound to an $\uparrow$-spin, and this bound state exchanges with singlet dimers
 in such a way that the phase strings signs cancel, as we will discuss in more detail in the next paragraphs.
 \item For a generic value of $p_\downarrow$, the phases $S_{p,p'}$ mismatch severely for a general off-diagonal path pair $(p,p')$. The exponent thereby rapidly oscillates as the
 path pair is varied, with the effect that their contributions cancel out. However, there are exceptions: the path $p$ may self-intersect forming closed loops;
 the path $p'$ then follows the same route, but either passing  the loops  in a different order [see, e.g., Fig.~\ref{Fig:interference}(b), top] or  passing the
 same loop along opposite directions [see, e.g., Fig.~\ref{Fig:interference}(b), bottom]. Although such two paths are off-diagonal, they have the same phase because of
 their identical lengths and thereby constructively interfere with each other. This interference picture is exactly the same as the one that underpins
Anderson localization [see Ref.~\cite{Tian04} for example]. { This indicates that the self-localization of the injected hole observed in  the numerical simulations as reported
first in Ref.~\cite{ZZ2013} has the same interference picture as familiar Anderson localization, provided that the conjecture of phase-string averages is valid. However, we should stress that to mobilize the canonical analytical theories developed for Anderson localization is a much more difficult task.}
\end{itemize}

\subsection{The single-hole momentum distribution}
\label{sec:momentum_peak}

Different from the standard Anderson localization mechanism in quenched disordered systems, our numerical simulations reported in the previous section
revealed that the phase-string driven self-localization mechanism goes hand in hand with a rather precise quantization of the states in single-hole
momentum space, i.e., the sharp peaks in the single-hole momentum distribution. The fact that the system retains perfect translational invariance is
of course a necessary condition for this to happen. However, it is far from obvious that the single-particle momentum can survive the severe ``annealed disorder"
associated with the spin fluctuations ``seeding" the phase-string interference. In addition, it has to be explained
why these single-hole wave vectors are strongly dependent on $\alpha$, clearly a parameter controlling the spin fluctuations in the quantum spin background.

It turns out that these behaviors are an immediate consequence of our assumption that the phase strings can be treated in a statistical way. Consider again
the single-hole Green's function (\ref{eq:20}) and set the coordinates $i=(x_i,0),j=(x_j,0)$. Take
$p_\downarrow=1$ such that  all paths have a positive sign, and it follows directly that  $G(j,i;E) \sim e^{i\pi (x_j-x_i)}$: the delocalizing hole
behaves as a quasiparticle with ground state  momentum $k_x=\pi$. This agrees with the numerical results in this regime, see the data corresponding to $\alpha=2/5$ in Fig.~\ref{Fig:Scaling}(a).

Taking generic values of $p_\downarrow < 1$ such that  self-localization occurs, the exponent in Eq.~(\ref{eq:20}) oscillates rapidly as the path is slightly deformed
and these contributions will in general cancel out. Therefore, the sum is dominated by the paths with a stationary phase. Varying the phase  $\pi p_\downarrow\int_p ds$,
we find that the resulting geodesic is a straight line connecting $i$ and $j$. Therefore,
\begin{eqnarray}
G(j,i;E)\sim e^{i\pi p_\downarrow (x_j-x_i)}.
\label{eq:19}
\end{eqnarray}
This implies a peak at $k_x=\pi p_\downarrow$ in the momentum distribution. Next, we observe that Eq.~(\ref{eq:17}) is not changed upon replacing $S_p$ by $-S_p$
since $e^{-2iS_p}=e^{-2\pi iN_p^\downarrow}=1$. Repeating the calculation for this choice, it follows that
\begin{eqnarray}
G(j,i;E)\sim e^{-i\pi p_\downarrow (x_j-x_i)}.
\label{eq:21}
\end{eqnarray}
This implies another peak in the momentum distribution, now at $k_x=2\pi-\pi p_\downarrow$.  This explains why the momentum distribution exhibits two peaks
at $k_x=\pi\pm \kappa$ according to the simulations in the self-localized regime [Fig.~\ref{Fig:Scaling}(a)]. This also sheds light on the mysterious dependence
on the rung coupling $\alpha$ of
the degree of the splitting of the momentum distribution into two peaks. According to our path-integral consideration,
\begin{equation}\label{eq:23}
    \kappa=\pi(1-p_\downarrow)\equiv \pi p_\uparrow,
\end{equation}
with $p_\uparrow$ being the probability for the hole to exchange its position with an $\uparrow$-spin nearest neighbor. The rung coupling is just tuning the number
of ``spin fluctuations" $p_\downarrow$ as of relevance to the ``dynamical phasing" of the hole wave function.

\begin{figure}[tbp]
\centerline{
    \includegraphics[height=2.3in,width=3.8in] {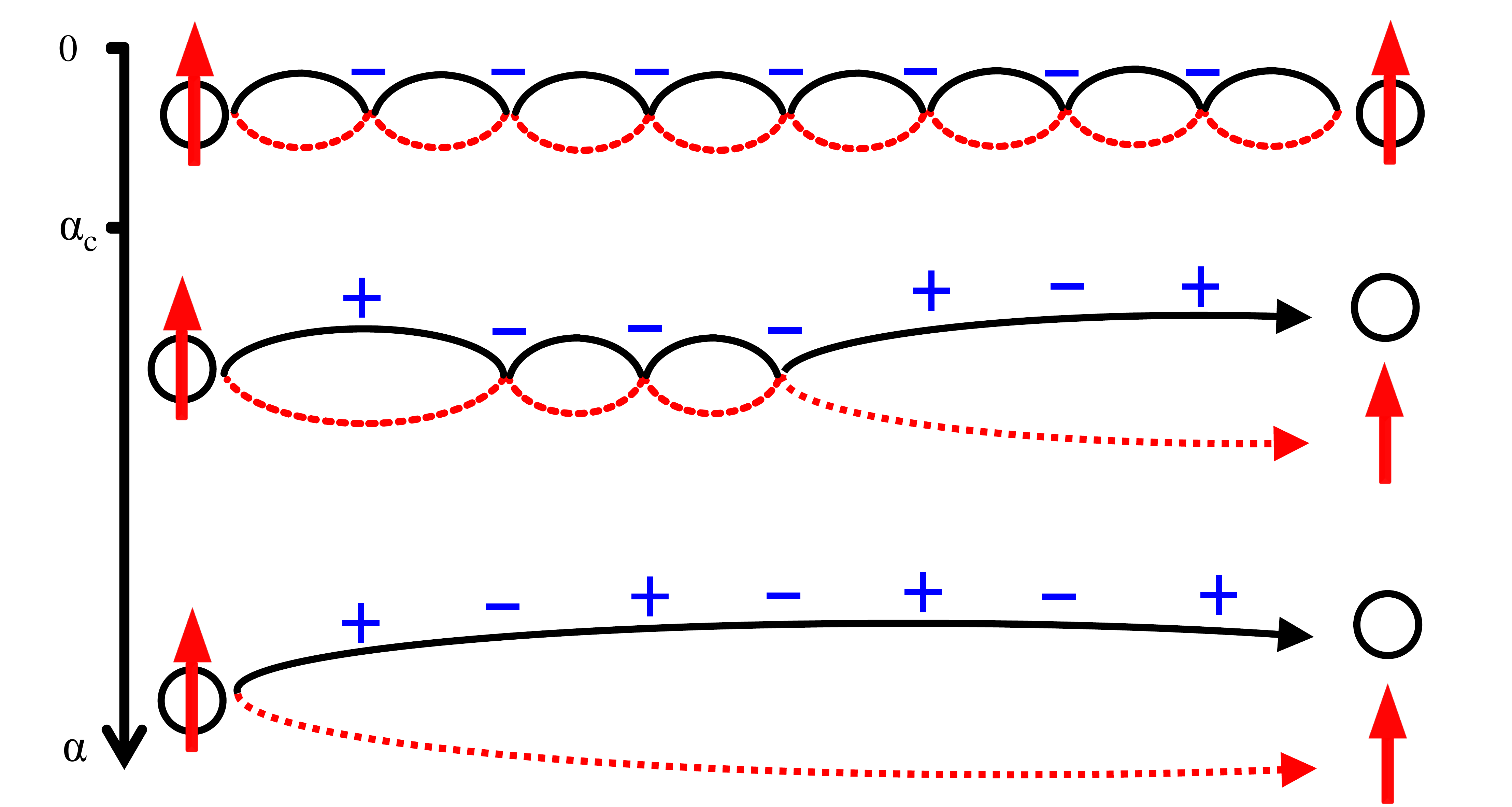}
    }
\caption{(Color online) Upon injecting a hole an $\uparrow$-spin is left. The motions of the hole and the $\uparrow$-spin are represented by the solid and dashed lines, respectively. The signs, $+/-$, are the bookkeeping of the exchange of the hole and the $\uparrow/\downarrow$-spins. Top: for $\alpha<\alpha_c$ the hole is always bound to the $\uparrow$-spin and they move together exchanging their position with spin singlets (nodes). Middle: for $\alpha>\alpha_c$ many spin singlets are broken. The hole and $\uparrow$-spin move separately and recombine whenever exchange their positions simultaneously with spin singlets. Bottom: for $\alpha\rightarrow\infty$ they move separately in the entire course of time.}
\label{Fig:Spin-Charge_Separation}
\end{figure}

For $\alpha<\alpha_c$ the spin pair living on the same rung forms a singlet state. The hole is
therefore always bound to an $\uparrow$-spin (represented by the nodes in Fig.~\ref{Fig:Spin-Charge_Separation})
since a spin singlet is broken and an $\uparrow$-spin is left after removing the $\downarrow$-spin electron (the realization of the single hole in the
simulations).   Subsequently, this hole-$\uparrow$-spin bound state exchanges its position with the spin singlet in the nearest rung, moving in the ladder
as a quasiparticle (cf. Fig.~\ref{Fig:Spin-Charge_Separation}, top). In more detail, such an exchange is a two-step process:  A hole-$\downarrow$-spin exchange  is followed by
the recombination of $\downarrow$- and $\uparrow$-spins into the singlet. The former step gives rise to a sign of $e^{i\pi}=-1$.
Because the hole -- bound with an $\uparrow$-spin -- could exchange its position only with $\downarrow$-spins, $p_\uparrow=0$
and therefore $\kappa=0$, such that  $k_x=\pi$. The phase string signs cancel out in this background formed by
the``strongly bound" valence bond solid and the hole just turns into a
quasiparticle which can be continued all the way in principle to a carrier living in an equivalent non-interacting band insulator.

When the rung coupling exceeds $\alpha_c$, the spin singlets in the background start to break up.
The injected hole and its ``partner" $\uparrow$-spin can now move independently.  These can recombine again upon exchanging their
positions simultaneously with a spin singlet (cf. Fig.~\ref{Fig:Spin-Charge_Separation}, middle). Importantly, because many singlets
are broken the hole has a non-vanishing probability to exchange
its position with $\uparrow$-spins. This probability $p_{\uparrow}$ increases with $\alpha$, to eventually saturate at $1/2$ when the motions of the hole
 and the $\uparrow$-spin are fully uncorrelated (cf. Fig.~\ref{Fig:Spin-Charge_Separation}, bottom). As a result, $\kappa$ increases with $\alpha$, to acquire a
 limiting value of  $\pi/2$, as found in the simulations [cf. Fig.~\ref{Fig:Scaling}(a)].  We expect on general grounds that in the limit of $\alpha\rightarrow\infty$
 the two-leg ladders should behave essentially in the same way as the  $t-J$ model model defined on a strictly one dimensional chain.
 It is a well-known result from Luttinger liquid theory that the single hole acquires a momentum $\kappa=\pi/2$ in this spin-charge separated case.

\subsection{The charge modulation}
\label{sec:charge_modulation}

We have collected all the pieces of the puzzle to explain the charge modulations observed in the DMRG simulations. On the one hand we have just learned
that due to the annealed phase-string disorder the hole can still propagate coherently at short distances, characterized by sharp single-particle
momenta ($\pi \pm \kappa$), where $\kappa$ has now an unconventional dynamical origin. However, at longer distances the phase-string ``enhanced backscattering"
interferences accumulate, resulting in the effect of the self-localization of the hole.
Suppose that the hole wave moves right along the $x$-direction, as described by $ \sim e^{i(\pi-\kappa) x}$
where the irrelevant amplitude is ignored and
($\pi-\kappa$) is the hole's momentum.
Next, for a confined motion to occur eventually the right-moving wave
must necessarily be reflected, where the scattering arises from those phase string fluctuations that mimic a disorder potential  .
The reflected momentum has to be ($\pi+\kappa$) since the momentum is quantized.
The reflected wave has the general form,
$\tilde r e^{i(\pi+\kappa) x}$, where $\tilde r$  is the complex reflection amplitude.
As a result, the hole density becomes
\begin{eqnarray}\label{eq:1}
n^h(x)&\sim& |e^{i(\pi-\kappa) x}+\tilde r e^{i(\pi+\kappa) x}|^2\nonumber\\
&=&(1+|\tilde r|^2) - 2|\tilde r| \sin(2\kappa x+\delta),
\end{eqnarray}
where $\tilde r \equiv i|\tilde r|e^{i\delta}$.
This shows that the self-localization inevitably leads to an interference pattern composed of a single harmonic, with a wave number given by
\begin{equation}\label{eq:24}
    Q_0=2\kappa,
\end{equation}
in agreement with our numerical findings summarized in Figs.~\ref{Fig:Q0} and \ref{Fig:Scaling}(b).

In the long wavelength limit the hole undergoes many of the aforementioned scattering events.
These events are independent and so are their scattering phases. Correspondingly, the oscillatory factor
in Eq.~(\ref{eq:1}) is replaced by $\sin(2\kappa x+\sum_{i=1}^n \delta_i)$,
where $i$ denotes the scattering event, $\delta_i$ is the scattering phase, and $n$ is the number of scattering events.
For large $n$ one may apply the law of large numbers to the modified factor, obtaining $\sin(2\kappa x)e^{-(\overline{\delta^2}-(\overline{\delta}^2)) n/2}$.
The exponent implies that the amplitude of the charge modulation exponentially decays to zero sufficiently far away from the center,
consistent with our DMRG simulations [cf. Fig.~\ref{Fig:HoleDensity}(a)].
Since at each scattering event the momentum state $(\pi\pm\kappa)$ is scattered into $(\pi\mp\kappa)$,
the momentum distribution peaks do not need to be associated with quasiparticles.
In addition, the self-localization may be considered as a consequence of multiple hole scattering.

{ We emphasize that the charge modulation cannot emerge from the coupling between
spin-exciton excitation and the moving holes. Indeed, the spin-exciton (polaron) leads to similar effects for both $t-J$ and $\sigma\cdot t-J$ models since these two models have the same spin gapped background. However, the charge modulation disappears totally for the latter model.}

\subsection{The oscillation of the ground-state energy in ladder's length}
\label{sec:relation}

For completeness, let us finalize this section by the analysis of a finite-size effect observed numerically, see Fig.~\ref{Fig:Edifference} [as well as Fig.~\ref{Fig:Scaling}(b)]. Specifically, we compactify the ladder into a ``ribbon'' by gluing its two ends together [cf. Fig.~\ref{Fig:Edifference}(a), upper right], turning the system into a ring with a
 circumference $N_x$. The ground-state energy of this compactified ladder consists of one part associated with the quantum spin system
 and the other part associated with the hole. To study the latter we note that the hole
 is already self-localized within the ring. Following the canonical treatise of the finite size effects on the usual
Anderson localization  in 1D \cite{Freilikher04} we may view such self-localization -- within the ring -- as free propagation confined by tunneling barriers placed
on both ends. Because of the aforementioned compactification the hole may circulate in the ring: this is equivalent to the motion in a $1$D
``perfect crystal'' with a (spatial) period of $N_x$ [Fig.~\ref{Fig:transmission}(a)], where the ``unit cells'' are created by periodically placed tunneling barriers.
 The energy associated with this crystal will have a tight-binding like band structure. The general form of the hole wave function will be   $
 \sim\cos(q_x N_x)$ where $q_x$ is the effective Bloch momentum.

Let a magnetic flux $\Phi$ pierce through the ring. The ground-state energy of the ring will depend on $\Phi$ by an amount $E_G^{{\rm 1-hole}}(\Phi)$.
Since the flux couples exclusively to the charge degree of freedom through $q_x \rightarrow q_x + \Phi/N_x$, leaving the spin system unaffected,
the flux dependence of the energy directly measures the hole contribution:  $\Delta E_{G}^{\text{1-hole}}(\Phi)\equiv E_{G}^{\text{1-hole}}(\Phi)-E_{G}^{\text{1-hole}}(0)$.
For $\Phi=\pi$ it follows that $\Delta E_{G}^{\text{1-hole}}(\pi)
\sim\cos (q_x N_x)$, while it also can be shown that (see Appendix~\ref{sec:band} for the proof),
\begin{equation}\label{eq:27}
    \cos(q_x N_x)
    \propto\cos (\kappa N_x + \delta).
\end{equation}
Note that $\delta$ is independent of $N_x$.
Taking this into account we obtain
\begin{equation}
 \Delta E_{G}^{\text{1-hole}}(N_x)\sim
 \cos (\kappa N_x+\delta).
 \label{eq:10}
 \end{equation}
This shows that $\Delta E_{G}^{\text{1-hole}}(N_x)$ oscillates in $N_x$, with the wave number
\begin{equation}\label{eq:25}
\kappa=\kappa_0.
\end{equation}
In combination with Eq.~(\ref{eq:24}) this yields
\begin{equation}\label{eq:26}
    Q_0=2\kappa=2\kappa_0,
\end{equation}
in agreement with the numerical finding [Fig.~\ref{Fig:Scaling}(b)].

\section{Discussion and outlook}
\label{sec:discussion}

Viewed from a general physics perspective, the considerations in this work have unveiled a new mechanism causing charge modulation associated with strongly interacting
fermions at a high density -- it is physics associated with the ``fermion  sign" problem. From this perspective, it is a cousin of the well-known Friedel oscillations of
the free fermion gas. In the free case, the ``fermion  signs" just translate to the Pauli principle having the ramification that the states reacting to an impurity
potential or an open boundary are quantum mechanical waves with a short wavelength inversely proportional to the Fermi momentum --
this period exhibits itself in the  conventional ``wiggles".
When Mott-ness takes over for sufficiently strong Coulomb repulsions the basic rules of the fermion statistics is drastically altered:
 because of the ``stay-at-home" principle the fermions turn
into spins at half filling, and spins are distinguishable particles. But in the presence of holes, fermion exchanges are restored in the ``intermediate vicinity" of the
delocalizing holes: the phase strings just enumerate how fermion signs re-enter the problem \cite{Zaanen08,Zaanen11,weng_07}.

In the single-hole case this boils down to the master rule that world histories in the path-integral formulation acquire an overall sign associated with the number of
times that the hole exchanges with a down-spin. As we demonstrated in section \ref{sec:interference}, for a particular path one can adopt the mean-field like approximation of averaging over these ``dynamical
signs". { Strikingly, with this approximation, a quantum mechanical like dynamics automatically emerges. For this ``emergent quantum mechanics'' a disordered potential is self-generated by the fluctuating spin background, and the phase arises from phase strings. By fundamental principles of physics of quantum disordered systems one may naturally expect the occurrence of Anderson localization like phenomenon. Indeed, we have shown an interference picture for self-localization which is the same as that for familiar Anderson localization. Of course, whether the quantitative behavior, such as the single parameter scaling \cite{Anderson79}, exists here calls for more sophisticated investigations.} The charge modulations occur automatically in this
 self-localized regime.  The averaging of phase strings makes it possible for the hole to propagate in a quantum mechanical fashion,
 characterized by wave vectors that are
 determined by the spin fluctuations. These in turn have in common with the usual Friedel oscillations
that an effective finite volume is needed for them to form standing waves corresponding to the charge modulations; otherwise the hole would
spread out over the entire system and the charge modulations would diminish.
However, completely different from the canonical quantum mechanics behind the Friedel oscillations, the modulation period is not at all set by density via the Fermi momentum,
but instead by the severity of the quantum spin fluctuations in the vacuum.

We emphasize that the mechanism is anchored in general principle and it should be very robust: By tailoring the right conditions it should be observable in the
laboratory. Ideally one would like to depart from an experimental realization of a two-leg ladder system characterized by a rung coupling $\alpha$ which
is strong enough. By electrolyte gating one could then force in holes
keeping the system very clean. The interest would be in the regime of very low carrier density, avoiding a metallic like behavior. The reason is that the long-ranged
Coulomb interaction should be kept strong enough to prohibit the pairing tendencies. Ideally a Wigner crystal would be realized; in the vicinity of the carriers one
would then expect our charge modulations which are in principle measurable by scanning tunneling spectroscopy. Their modulation wavelength could then be
related to the properties of the spin system by carefully quantifying the latter by inelastic neutron scattering and so forth.

We have analyzed here specifically the case of a two-leg ladder doped with a single hole. What are the ramifications in general, considering
lattices of higher dimensions or finite hole densities?
We are exploring these at present, and let us present some preliminary results.
For the single-hole doped case, the charge modulation has been found to persist in other even-leg $t-J$ ladders as well,
 provided that a spin gap is present in the spin system.
For an odd-leg ladder, we do not find the charge modulation even though the charge remains localized. This appears to be related  to
a particular self-averaging of the scattering phase $\delta$ caused by the gapless spin fluctuations, smearing the interference patterns
associated with the left- and right-moving propagating modes.
The charge modulation is expected to reemerge in a gapless spin system only at finite doping, when the spin-spin correlation
in the background becomes short-ranged due to the doping.

Given that we are dealing with a strongly interacting, dense fermion system it is a-priori not obvious whether the physics of a single carrier in isolation
has dealings with the physics of the system at any finite density. One immediate ``complication" is by now well established:
 Two holes in a $t-J$ ladder
will immediately bind in a ``Cooper pair", where yet again the phase strings play a crucial role in the binding mechanism \cite{ZZ2014}.
Turning to the truly finite density case on many-rung ladders
(or the 2D square lattice) the DMRG computations already showed a long time ago the 214-like stripes \cite{White98}. These appear to be formed in essence from the
preformed pairs and these are not supposed to show the single hole modulations, as we just argued. However, we have preliminary indications that
the phase strings also control the truly co-operative translational symmetry breaking in this case, although the precise mechanism appears to be different.
Upon switching off the phase strings by employing the $\sigma\cdot t-J$ model, we find that the stripes do disappear. We leave a precise analysis of this physics
to a future publication.

We wish to mention that discussions on the weak coupling two-leg Hubbard ladders have been well documented where, notably, a standard quasiparticle picture works well \cite{Scalapino96,Ludwig01}. An outstanding problem left by the present work is, i.e., how to connect such a weak coupling regime with the strong coupling regime, especially in the parameter regime where the half-filled ground state remains fully gapped in both weak and strong coupling. Based on our present results, we expect that the weak and strong coupling limits might be smoothly connected in the strong rung regime, where the quasiparticle behavior is well established at $\alpha<\alpha_c$ in the $t-J$ ladder. Whereas at $\alpha>\alpha_c$, a ``phase transition'' in Mott nature is expected so that the quasiparticle behavior for weak coupling could be replaced by the charge localization for strong coupling observed in this work.
This issue is currently under investigations.

\subsection{Comments on a followup work}
\label{sec:comments}

After this paper was submitted, a DMRG paper has recently appeared in Ref.~\onlinecite{White2015}, where critical comments on the present work were made. First of all, the numerical simulations in that work confirm some of the main results reported here and in an earlier work \cite{ZZ2014qp}, namely the charge modulation and the existence of a quantum critical point in the anisotropic $t$-$J$ ladders. However, the authors of that paper made a drastically different interpretation about the physics happening at $\alpha>\alpha_c$. They claim that the quasiparticle picture is still valid accompanying the momentum splitting shown in Fig.~\ref{Fig:Scaling}(a) at $\alpha>\alpha_c$, in contrast to the quasiparticle collapsing and localization picture discussed here. Their arguments are based mainly on the numerical finding of the finiteness of $Z=\sum_i|\langle \phi|c_{i\downarrow}^\dagger|\Psi\rangle|^2$ on both sides of $\alpha_c$ (here $|\phi\rangle$ and $|\Psi\rangle$ denote the undoped and one-hole ground states, respectively).  But we point out that a finite quasiparticle spectral weight should be associated with the quantity $Z_k=|\langle \phi|c_{k\downarrow}^\dagger|\Psi\rangle|^2$, not $Z$. For a translationally invariant system, these two quantities are indeed equivalent since the ground state has a well-defined Bloch momentum. But in the presence of localization and accompanied charge modulation, the translational symmetry is broken. Hence, one cannot validate the quasiparticle picture by a finite $Z$ because it may be contributed by a spectrum of momenta in the ground state (recalling $Z=\sum_k Z_k$). As a matter of fact, in Fig.~\ref{Fig:Edifference} we have compared the ground-state energy difference between the periodic and antiperiodic boundary conditions imposed on the charge sector, and the envelope of this energy deviation (as a function of ladder length $N_x$) decays exponentially with $N_x$. This Thouless-like criterion for Anderson localization is a direct characteristic of the hole localization. In particular, the energy difference oscillates with $N_x$ with a characteristic momentum precisely locked with the half of $Q_0$ for the charge modulation [Fig.~\ref{Fig:Scaling}(b)], which has been also naturally explained based on the localization picture in Sec. III D. As one more step, we have further examined the $\sigma\cdot t$-$J$ ladder which differs from the $t$-$J$ ladder solely by switching off the phase string effect.  There, the quasiparticle behavior is recovered: Both charge modulation and localization disappear together with the quantum critical point $\alpha_c$. This clearly established the connection between the phase string effect and the new phase at $\alpha>\alpha_c$.

Finally, we wish to point out that whether the charge modulation is due to the quantum interference of the phase string effect discussed here or is simply attributed to a standing wave of the quasiparticle as suggested in Ref.~\cite{White2015} has far reaching different physical implications. The authors in Ref.~\cite{White2015} have introduced a phenomenological model claimed to enable adiabatically connecting two phases with a noninteracting band model. However, it has been shown \cite{ZZ2014qp} that the two-hole binding energy becomes finite and substantial in the regime of $\alpha > \alpha_c$. This would be contradictory to the above quasiparticle picture, where the two-hole state would be adiabatically connected to an unpaired state of the band model given in Ref.~\cite{White2015}.

\section*{Acknowledgements}
\label{sec:acknowledgements}

We acknowledge discussions with L. Fu, T. L. Ho, D. H. Lee, J. W. Mei, D. N . Sheng, and X. G. Wen. C.T. would like to thank Yu. P. Bliokh and V. D. Freilikher for teaching him the resonator model of Anderson localization. Work done at Tsinghua Univ. was supported by the NSFC (Nos. 11174174 and 11104154), by the NBRC (973 Program, Nos. 2010CB923003 and 2011CBA00108), and by the Tsinghua Univ. ISRP. Work by H.C.J was supported by the Department of Energy,
Office of Science, Basic Energy Sciences, Materials Sciences and Engineering Division, under Contract DE-AC02-76SF00515. J.Z. acknowledges support of a grant from the John Templeton foundation. The opinions expressed in this publication are those of the authors and do not necessarily reflect the views of the John Templeton foundation.

\appendix

\section{Proof of Eq.~(\ref{eq:27})}
\label{sec:band}
\renewcommand{\thefigure}{A\arabic{figure}}
\setcounter{figure}{0}
\renewcommand{\theequation}{A\arabic{equation}}
\setcounter{equation}{0}

\begin{figure}[tbp]
\centerline{\includegraphics[height=2.7in,width=3.2in] {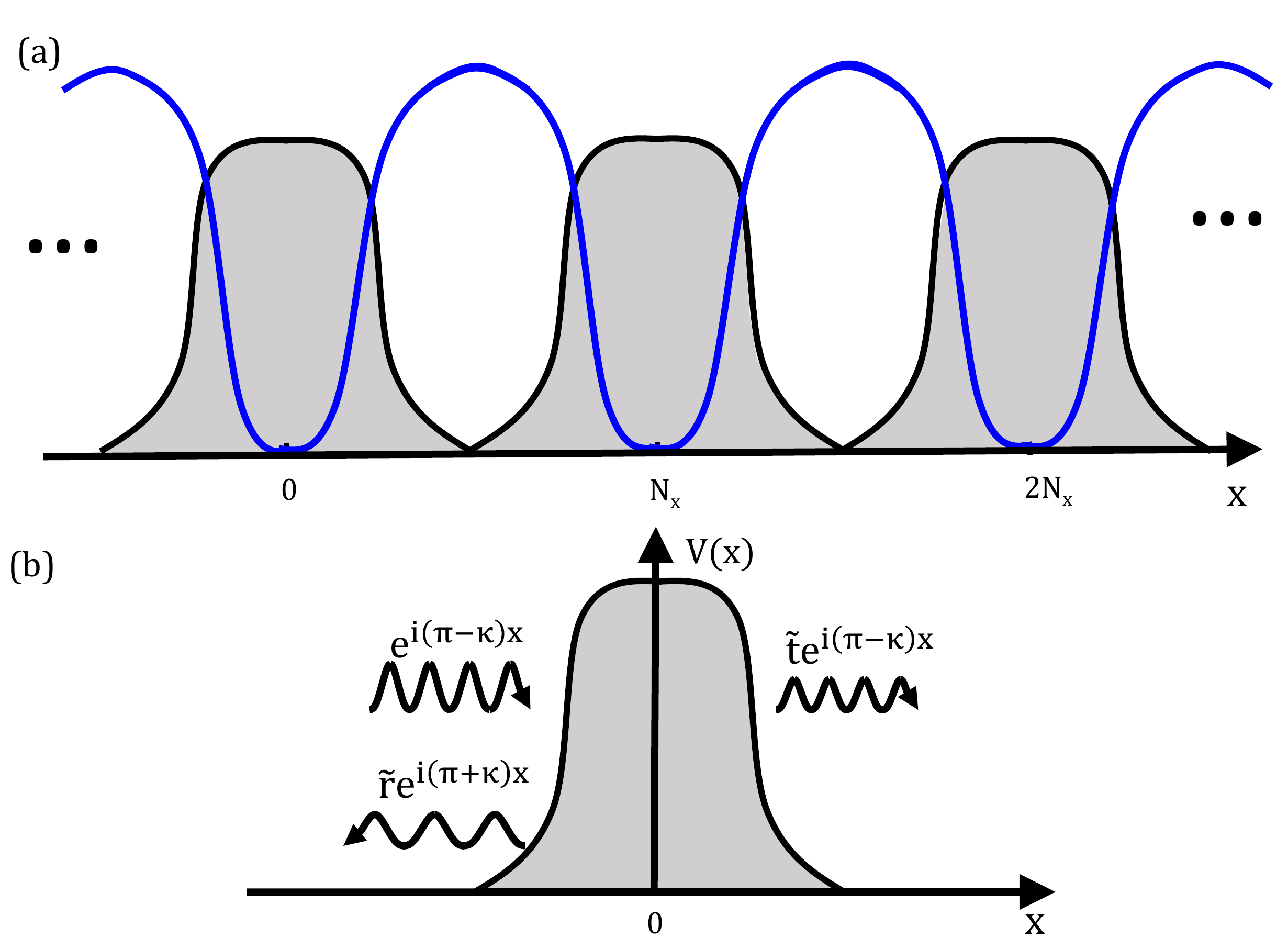}}
\caption{(a) By enforcing the periodic boundary condition on a ladder we obtain an infinite periodic system. The period of this $1$D ``crystal'' is $N_x$. The self-localized profile of the hole density (blue solid curve) within a single unit cell implies that the unit cells are weakly coupled to each other via a symmetric tunneling barrier (grey area). (b) A plane wave of momentum $(\pi-\kappa)$ incident on a single tunneling barrier is partially transmitted and reflected. The reflected wave has a momentum of $(\pi+\kappa)$.}
\label{Fig:transmission}
\end{figure}

Similar to Anderson localization in $1$D disordered systems \cite{Freilikher04}, the self-localization of the hole injected into a $t-J$ ladder may be viewed as a quantum motion, albeit arising from phase strings, confined by two tunneling barriers. When the ladder is compactified by a periodic boundary condition, the tunneling barrier, denoted as $V(x)$, is replicated infinite times and periodically positioned in the $x$-direction: effectively, we obtain a $1$D perfect ``crystal'' with the ``unit cell" of size $N_x$ [Fig.~\ref{Fig:transmission}(a)]. In this case the hole's energy must form a band. Below we largely follow the method of Ref.~\cite{Mermin} to study this band.

Suppose that only one tunneling barrier is present, which is assumed to be located at $x=0$ without loss of generality, and that a plane wave of momentum $(\pi-\kappa)$ is incident on the barrier from the left [Fig.~\ref{Fig:transmission}(b)]. The scattered wave must have the form
\begin{eqnarray}
\label{eq:11}
  \psi_L(x) = \bigg\{\begin{array}{c}
                       e^{i(\pi-\kappa)x} + \tilde r e^{i(\pi+\kappa)x}, \quad x\leq -\frac{N_x}{2},\\
                       \tilde t e^{i(\pi-\kappa)x}, \quad x \geq \frac{N_x}{2},\,\,\,\, \quad\quad\quad\quad\quad\quad
                     \end{array}
\end{eqnarray}
where $\tilde r$ and $\tilde t$ are the (complex) reflection and transmission amplitudes, respectively. Since the tunneling barrier is symmetric with respect to its center, the following scattered wave,
\begin{eqnarray}
\label{eq:12}
  \psi_R(x) = \bigg\{\begin{array}{c}
                       \tilde t e^{i(\pi+\kappa)x}, \quad x \leq -\frac{N_x}{2},\,\,\,\,\, \quad\quad\quad\quad\quad\quad\\
                       e^{i(\pi+\kappa)x} + \tilde r e^{i(\pi-\kappa)x}, \quad x\geq \frac{N_x}{2}\quad\quad\,
                     \end{array}
\end{eqnarray}
has the same energy [indeed, $\psi_R(-x)=\psi_L(x)$]. In the scattering region, $|x|\leq \frac{N_x}{2}$,
the wave function $\psi(x)$ must be the superposition of $\psi_{L,R}(x)$.

Since the Hamiltonian of the crystal in the region of $|x|\leq \frac{N_x}{2}$ is identical to that with single tunneling barrier, the wave function in this region must be the same as $\psi(x)$. Then, by the Bloch's theorem we find $\psi(\frac{N_x}{2})=e^{iq_xN_x}\psi(-\frac{N_x}{2})$ and $\frac{d}{dx}\psi(x)|_{x=\frac{N_x}{2}}=e^{iq_xN_x}\frac{d}{dx}\psi(x)|_{x=-\frac{N_x}{2}}$. Recall that $q_x$ is the Bloch momentum. Substituting Eqs.~(\ref{eq:11}) and (\ref{eq:12}) into these two relations we obtain
\begin{equation}\label{eq:13}
    \cos(q_x N_x)=\frac{\tilde t^2-\tilde r^2}{2\tilde t}e^{i\kappa N_x}+\frac{1}{2\tilde t}e^{-i\kappa N_x}.
\end{equation}
For the self-localization, $|\tilde t|
$ is exponentially small for $N_x$ much larger than the localization length. Because of this $|\tilde t|\ll |\tilde r|\approx 1$, and Eq. ~(\ref{eq:13}) is simplified to
\begin{equation}\label{eq:14}
    \cos(q_x N_x)=\frac{1}{2\tilde t}\left(-\tilde r^2 e^{i\kappa N_x} + e^{-i\kappa N_x}\right).
\end{equation}
Since the left-hand side is real, $\tilde t$ and $\tilde r$ must take the general forms of $\tilde t=|\tilde t|e^{i\delta}$ and $\tilde r=i|\tilde r|e^{i\delta}$. Substituting them into Eq.~(\ref{eq:14}) gives Eq.~(\ref{eq:27}).


\begin{thebibliography}{99}

\bibitem{Zaanen89} J. Zaanen and O. Gunnarsson, Phys. Rev. B \textbf{40}, 7391(R) (1989).

\bibitem{Machida89} K. Machida, Physica C \textbf{158}, 192 (1989).

\bibitem{Uchida95} J. M. Tranquada, B. J. Sternlieb, J. D. Axe, Y. Nakamura, and S. Uchida, Nature \textbf{375}, 561 (1995).

\bibitem{Zaanen10} J. Zaanen, {\it 100 Years of Superconductivity},  edited by  H. Rochalla and P.H. Kes, (Chapman and Hall, London, in press, 2015).

\bibitem{Zaanen14} B. Keimer, S. A. Kivelson, M. R. Norman, S. Uchida, J. Zaanen, arXiv: 1409.4673, and the references therein.

\bibitem{Sachdev14d} L. E. Hayward, D. G. Hawthorn, R. G. Melko, and S. Sachdev, Science \textbf{343}, 1336 (2014).

\bibitem{Efetov14a} H. Meier, C. P$\acute{\rm e}$pin, M. Einenkel, and K. B. Efetov, Phys. Rev. B, \textbf{89}, 195115 (2014).

\bibitem{Mei14} L. Zhang and J. W. Mei, arXiv:1408.6592, and the references therein.

\bibitem{White98} S. R. White and D. J. Scalapino, Phys. Rev. Lett. \textbf{80}, 1272 (1998).

\bibitem{Troyer14} P. Corboz, T. M. Rice, and M. Troyer, Phys. Rev. Lett. \textbf{113}, 046402 (2014).

\bibitem{Khomskii68} L. N. Bulaevskii, E. L. Nagaev, and D. L. Khomskii,
Zh. Eksp. Teor. Fiz. \textbf{54}, 1562 (1968) [Sov. Phys. JETP \textbf{27}, 836 (1968)].

\bibitem{Brinkman70}W. F. Brinkman and T. M. Rice, Phys. Rev. B \textbf{2}, 1324 (1970).

\bibitem{SCBA} S. Schmitt-Rink, C. M. Varma and A. E. Ruckenstein, Phys. Rev. Lett. \textbf{60}, 2793 (1988); C. L. Kane, P. A. Lee, and N. Read, Phys. Rev. B \textbf{39}, 6880 (1989); G. Martinez and P. Horsch, Phys. Rev. B \textbf{44}, 317 (1991).

\bibitem{Sheng1996} D. N. Sheng, Y. C. Chen, and Z. Y. Weng, Phys. Rev. Lett. \textbf{77}, 5102 (1996).

\bibitem{Weng1997} Z. Y. Weng, D. N. Sheng, Y. C. Chen, and C. S. Ting, Phys. Rev. B \textbf{55}, 3894 (1997).

\bibitem{weng_07} Z. Y. Weng, Int. J. Mod. Phys. B \textbf{21}, 773 (2007), and the references therein.

\bibitem{Zaanen08} K. Wu, Z. Y. Weng, and J. Zaanen, Phys. Rev. B \textbf{77}, 155102 (2008).

\bibitem{Zaanen11} J. Zaanen and B. J. Overbosch, 	Phil. Trans. R. Soc. A \textbf{369}, 1599 (2011).

\bibitem{ZZ2013} Z. Zhu, H. C. Jiang, Y. Qi, C. S. Tian and Z. Y. Weng, Sci. Rep. \textbf{3}, 2586 (2013).

\bibitem{Weng2001} Z. Y. Weng, V. N. Muthukumar, D. N. Sheng, and C. S. Ting, Phys. Rev. B \textbf{63}, 075102 (2001).

\bibitem{Lee85} {\it Fifty Years of Anderson Localization}, edited by E. Abrahams (World Scientific, Singapore, 2010).

\bibitem{ZZ2014qp} Z. Zhu and Z. Y. Weng, arXiv:1409.3241.

\bibitem{White92} S. R. White, Phys. Rev. Lett. \textbf{69}, 2863 (1992).

\bibitem{Dagotto99} E. Dagotto, Rep. Prog. Phys. \textbf{62}, 1525 (1999).

\bibitem{Tian04} C. Tian, A. Kamenev, and A. Larkin, Phys. Rev. B \textbf{72}, 045108 (2005).

\bibitem{Anderson79} E. Abrahams, P. W. Anderson, D. Licciardello, and T.V. Ramakrishnan, Phys.
Rev. Lett. \textbf{42}, 673 (1979).

\bibitem{ZZ2014} Z. Zhu, H. C. Jiang, D. N. Sheng and Z. Y. Weng, Sci. Rep.  \textbf{4}, 5419 (2014).

\bibitem{Scalapino96} H. Endres, R.M. Noack, W. Hanke, D. Poilblanc, D.J. Scalapino, Phys. Rev. B {\bf 53}, 5530 (1996).

\bibitem{Ludwig01} R. Konik and A. A. Ludwig, Phys. Rev. B \textbf{64}, 155112 (2001).

\bibitem{Freilikher04} K. Yu. Bliokh, Yu. P. Bliokh, and V. D. Freilikher,

J. Opt. Soc. Am. B \textbf{21}, 113 (2004); K. Yu. Bliokh, Yu. P. Bliokh, V. Freilikher, S. Savel'ev, and F. Nori, Rev. Mod. Phys.
\textbf{80}, 1201 (2008).

\bibitem{Mermin} N. W. Aschroft and N. D. Mermin, {\it Solid state physics} (Harcourt college publishers, Fort Worth, 1976), pp. 146.

\bibitem{White2015} S. R. White, D. J. Scalapino, and S. A. Kivelson, arXiv:1502.04403.


\end{thebibliography}
\end{document}